\documentclass{elsarticle}
\usepackage{natbib}
\usepackage{geometry}
\usepackage{hyperref}
\usepackage{graphicx} 
\usepackage{amsfonts}
\usepackage{amssymb}
\usepackage{mathtools}
\usepackage{mathrsfs}
\usepackage{txfonts}
\usepackage{dsfont}
\usepackage{pifont}
\usepackage{subcaption}
\usepackage{booktabs}
\usepackage{xcolor}
\usepackage{tabularx}
\usepackage{multirow}
\usepackage{multicol}
\usepackage{cleveref}
\usepackage{float}
\usepackage{arydshln}
\usepackage{tikz,pgfplots}
\pgfplotsset{compat=1.18}
\usepackage{bm}   
\usepackage{bbm}

\title{Multi-level Traffic-Responsive Tilt Camera Surveillance through Predictive Correlated Online Learning}
\author[1]{Tao Li\fnref{contrib}}
\ead{tl2636@nyu.edu}
\author[2]{Zilin Bian\fnref{contrib}\corref{poc}}
\ead{zb536@nyu.edu}
\author[1]{Haozhe Lei}
\ead{hl4155@nyu.edu}
\author[2]{Fan Zuo}
\ead{fan.zuo@nyu.edu}
\author[1]{Ya-Ting Yang}
\ead{yy4348@nyu.edu}
\author[1]{Quanyan Zhu}
\ead{qz494@nyu.edu}
\author[3]{Zhenning Li}
\ead{zhenningli@um.edu.mo}
\author[2]{Kaan Ozbay}
\ead{kaan.ozbay@nyu.edu}

\fntext[contrib]{Authors contributed equally}
\cortext[poc]{Corresponding author: Zilin Bian, email: zb536@nyu.edu}

\address[1]{Department of Electrical and Computer Engineering,  New York University}
\address[2]{Department of Civil and Urban Engineering, New York University}
\address[3]{State Key Laboratory of Internet of Things for Smart City, University of Macau}

\begin{document}
\begin{abstract}

In urban traffic management, the primary challenge of dynamically and efficiently monitoring traffic conditions is compounded by the insufficient utilization of thousands of surveillance cameras along the intelligent transportation system.  This paper introduces the multi-level Traffic-responsive Tilt Camera surveillance system (TTC-X), a novel framework designed for dynamic and efficient monitoring and management of traffic in urban networks. By leveraging widely deployed pan-tilt-cameras (PTCs), TTC-X overcomes the limitations of a fixed field of view in traditional surveillance systems by providing mobilized and 360-degree coverage. The innovation of TTC-X lies in the integration of advanced machine learning modules, including a detector-predictor-controller structure, with a novel Predictive Correlated Online Learning (PiCOL) methodology and the Spatial-Temporal Graph Predictor (STGP) for real-time traffic estimation and PTC control. The TTC-X is tested and evaluated under three experimental scenarios (e.g., maximum traffic flow capture, dynamic route planning, traffic state estimation) based on a simulation environment calibrated using real-world traffic data in Brooklyn, New York. The experimental results showed that TTC-X captured over 60\% total number of vehicles at the network level, dynamically adjusted its route recommendation in reaction to unexpected full-lane closure events, and reconstructed link-level traffic states with best MAE less than 1.25 vehicle/hour. Demonstrating scalability, cost-efficiency, and adaptability, TTC-X emerges as a powerful solution for urban traffic management in both cyber-physical and real-world environments.
\end{abstract}
\maketitle
\section{Introduction}
The intelligent transportation system (ITS) in urban areas has significantly expanded traffic surveillance cameras along road segments. These cameras are strategically placed at key locations such as road intersections, aiming to help traffic management agencies by providing real-time traffic information such as traffic congestion, traffic incidents, special events\cite{bian2019estimating,bian2021time}, and so on. Unlike traditional surveillance cameras with fixed fields of view (FOVs) that offer limited, single-direction \cite{akilan2019video} traffic monitoring, pan-tilt-cameras (PTCs) offer a more dynamic solution with their 360-degree mobilized coverage \cite{haghighat2023computer}, enabling versatile monitoring across different areas. Although the PTCs allow for more mobility and coverage, the widespread camera network (with PTCs) remains underutilized. PTCs operate in a static mode, oriented in a single direction at one time, and these PTCs within the network still function in isolation without inter-camera communication \cite{yang2023cooperative}. This leads to prolonged inactivity periods, during which traffic management agencies must manually reposition the cameras for specific monitoring tasks. Consequently, when traffic situations (e.g., congestion, accidents) arise along the road, the PTCs still require manual intervention to adjust their orientation appropriately for monitoring and management purposes. 
\begin{figure}[t!]
    \centering
    \includegraphics[width=0.85\columnwidth]{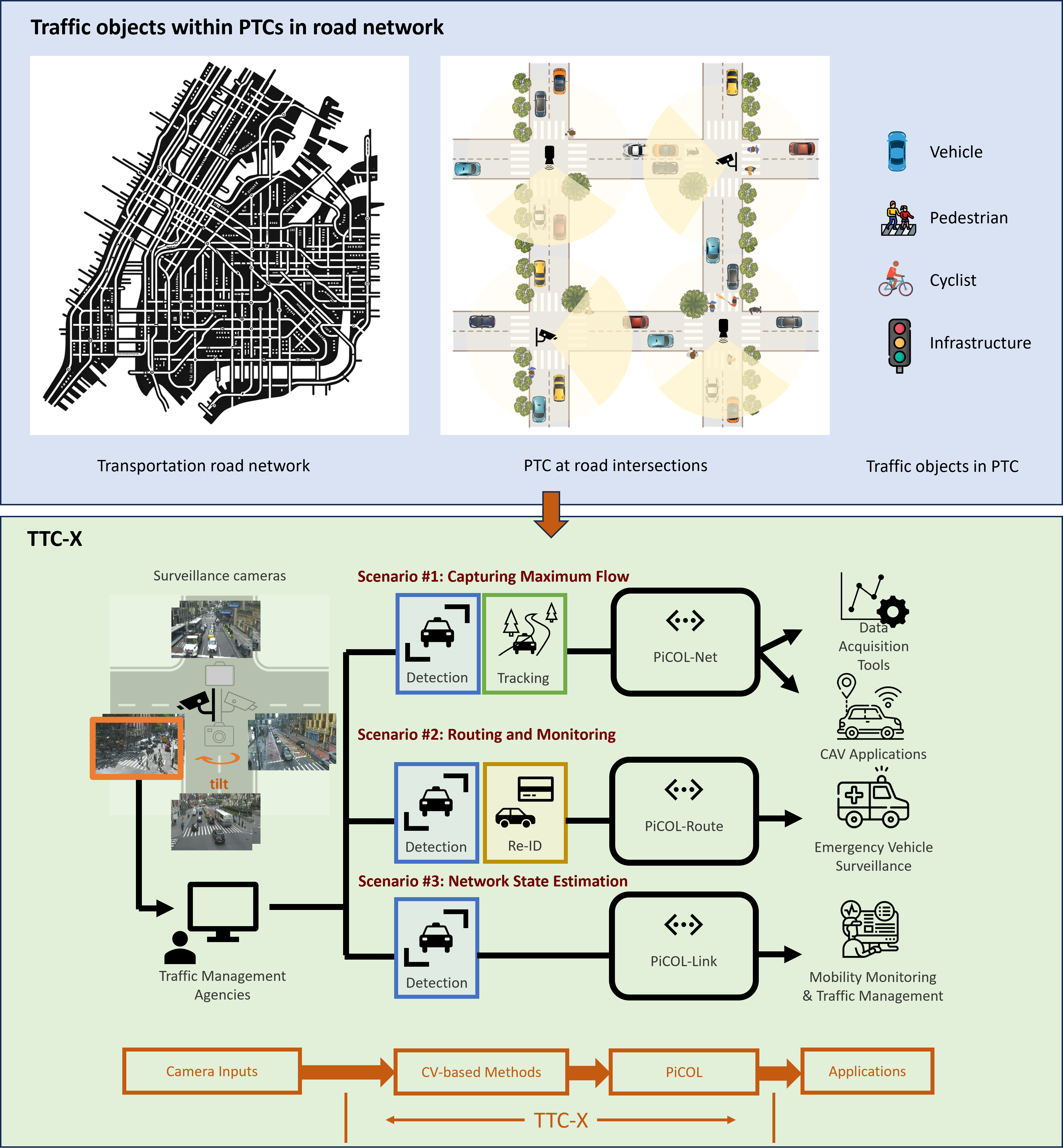}
    \caption{TTC-X conceptualization and formulation}
    \label{fig:ptc-network}
\end{figure}
One way to avoid the manual intervention of PTCs and labor-intensive costs from transportation management agencies is to design an automated and responsive traffic surveillance system. Seizing the opportunities of the current broadly deployed PTC system along intersections in urban areas, the idea is to extend from a single-camera-single-view to a multi-level surveillance strategy encompassing link, route, and network-level information. As shown in \Cref{fig:ptc-network}, each PTC covers four oriented directions, and each orientation covers one roadway segment with traffic from at most two directions. Therefore, the workflow of designing such a PTC-based traffic surveillance system includes 1) detecting and tracking multiple traffic objects (e.g., vehicle, pedestrian, cyclist, incident) from PTCs, 2) learning and estimating the mobility nature of traffic objects, and 3) collaboratively controlling multiple PTCs to achieve multi-level monitoring tasks for specific traffic object. However, this approach presents several challenges:
\begin{itemize}
    \item \textbf{Complexity and dynamics in urban transportation network:} Urban transportation networks are inherently complex and dynamic, characterized by intricate road connections and diverse network structures coupled with time-variant traffic demand and frequent traffic incidents. A challenging task is to seize traffic dynamics in both spatial and temporal dimensions in constraints to road network features to help the multi-camera surveillance system. 
    \item \textbf{Cost-efficient monitoring of large-scale transportation networks:} While PTCs offer extended coverage and mobility, effectively utilizing them to monitor large-scale networks with minimal equipment poses a significant challenge. For example, as shown in \Cref{fig:ptc-network}, on a theoretical basis, only four PTCs (one PTC at each intersection) are needed to monitor traffic along 24 road segments. However, the practical implementation faces the hurdle of coordinating these cameras to cover multiple directions simultaneously without losing focus on critical areas. This necessitates the development of intelligent algorithms that can dynamically prioritize camera orientations based on real-time traffic conditions and historical data trends.
    \item \textbf{Rapid and automatic response to traffic situations:} Given the complicated and dynamic nature of traffic patterns across a large-scale transportation network, it is essential for a surveillance system to proactively and responsively adapt to various traffic situations, such as congestion or incidents. For example, when a non-recurrent traffic incident occurs on a roadway segment, disrupting traffic flow, the disturbance evolves dynamically in both temporal and spatial dimensions. Detecting such changes in real-time and responding appropriately is a critical challenge, requiring advanced algorithms and responsive system architecture.
\end{itemize}

This study proposes a novel Traffic-responsive Tilt Camera surveillance system for everything (X) in a large-scale transportation network (TTC-X) to overcome the aforementioned challenges and hurdles. The TTC-X framework is innovatively designed based on a detector-predictor-controller (DPC) structure, integrating various technologies for comprehensive traffic monitoring. Specifically, the detector component utilizes established video extraction methods, including multiple-object-detection (MOD)\cite{redmon2016look}, tracking\cite{wojke2017simple,Tang_2019,neupane2022real}, and re-identification (Re-ID)\cite{shen2023triplet, zhou2022gan}. The appropriate use of the mentioned video extraction methods enables TTC-X to handle multiple objects captured by PTCs in large-scale transportation road networks. Theoretically, anything within the frame of PTC can be tracked and extracted into digital information, as shown in Fig.\ref{fig:ptc-network}. Multiple traffic objects, including vehicles, pedestrians, cyclists, and infrastructure, can be extracted using appropriate video extraction methods. 

To comprehensively demonstrate the proposed TTC-X, we limit our research scope and focus on vehicles as primary traffic objects. As shown in \Cref{fig:ptc-network}, this study presents three distinct scenarios captured by PTCs within the transportation network, each illustrating different levels of traffic information: network, route, and link level. Due to space constraints, these scenarios are selectively demonstrated to encompass the comprehensive capabilities of TTC-X.

At the network level, TTC-X is designed to capture and collect maximum traffic flow data across the entire transportation network in real-time. This scenario focuses on the overarching view of traffic dynamics, aiming to optimize the overall traffic flow and reduce congestion. With the help of appropriate MOD and tracking methods, TTC-X ensures the capture of the maximum number of vehicle trajectories, supporting applications such as data acquisition tools and connected and autonomous vehicle (CAV) simulation. In the route-level scenario, TTC-X facilitates real-time route planning. It identifies paths with the least travel time, considering unexpected traffic disruptions. This feature is handy for dynamic route optimization in response to evolving traffic conditions, such as accidents or road closures. Unlike probe-vehicle-based routing methods, this real-time route planning is high resolution and high accuracy thanks to the PTC surveillance cameras, which can provide reliable routing for cases requiring precise routing, such as emergency vehicle routing. The link-level scenario encompasses two sub-tasks: traffic state reconstruction and forecasting. The reconstruction task ensures that TTC-X accurately recovers the most realistic current traffic states across all road segments within the network. Meanwhile, the forecasting task enables TTC-X to provide confident predictions under various traffic situations, including sudden disturbances like non-recurrent traffic incidents.

The TTC-X framework integrates the Predictive Correlated Online Learning (PiCOL) methodology, which bridges the predictor and controller to provide optimal and responsive PTC tilting strategies. Within PiCOL, the Spatial-Temporal Graph Predictor (STGP) plays a crucial role. STGP is tasked with estimating and forecasting traffic states on each road segment within the transportation network. It operates with a plug-and-play online learning controller that distributedly calibrates each PTC's tilting strategy in correlation with the rest online. This correlated online learning leads to scalable, responsive, and coordinated PTC coverage over the road network in dynamic traffic patterns. The predictor and the controller function in real-time at two complementary levels. At the first level, STGP monitors and supplements traffic state information for road segments not directly observed by the PTCs whose tilting policy is dedicated by the controller. The controller, in turn, fuses these estimates with its observations to enhance the accuracy and reliability of STGP's predictions. At the second level, STGP's estimations consider the nature of traffic flow and network structural constraints. This enables STGP to confidently predict traffic states for segments the controller has not observed directly, particularly by considering observations from neighboring segments. Through this approach, STGP leverages network topology and local spatial dependencies to learn and infer traffic patterns, thereby enhancing the overall effectiveness of the TTC-X system in managing and responding to dynamic traffic conditions.

The performance of the TTC-X framework is evaluated in each of three scenarios using a simulation dataset calibrated using real-world data from Brooklyn, New York. At the network level, it is found that TTC-X can capture over 60\% number of vehicles within the road network throughout the day. With the help of STGP, TTC-X can estimate the traffic demand within the road network with MAPE (Mean Absolute Percentage Error) less than 10\%. At the route level, TTC-X is assessed using a case study with a full-lane closure event in the road network. TTC-X identified the optimal route within 5min and provided accurate trip travel time with an error of less than 1.6\%. At the link level, TTC-X achieved the best MAE (Mean Absolute Error) as 12.36 vehicles per hour for the forecasting task and 1.25 for the reconstruction task.

The contributions of this study are summarized below.
\begin{itemize}
    \item We propose a novel framework for the Traffic-responsive Tilt Camera surveillance (TTC-X) system, a versatile PTC-based surveillance system designed for network-wide, real-time traffic monitoring and management. TTC-X integrates advanced machine learning modules across its detector, predictor, and controller components, enabling real-time multi-camera coordination. The proposed TTC-X offers responsive intelligence that rapidly adapts PTC tilting on the fly to a variety of traffic situations. 
    \item A cutting-edge distributed PTC control framework, PiCOL is proposed for forecasting and reconstructing multi-level traffic information by integrating an offline learning-based predictive model named STGP into an online learning control paradigm. With the customization of loss functions, PiCOL is eligible to extend into applications driven by multiple tasks.
    \item We design three case scenarios at the network, route, and link levels, targeting different real-world application usage. Findings reveal that TTC-X can forecast and reconstruct multi-level (both network and link level) traffic information using a minimal set of PTCs with low information loss. Moreover, abrupt disturbances in the traffic network can be swiftly handled by TTC-X at both route- and link-level. The optimal route can be updated, and link-level traffic states can be estimated in reaction to abrupt changes with rapid responsiveness.
    \item The light-weighted, scalable, cost-efficient features of the TTC-X framework enables its connection with wide-range applications, in both cyber-physical world and real world environment. In the cyber-physical realm, TTC-X can be integrated within a virtual setup such as CARLA-SUMO co-simulation, enabling rigorous testing and refinement in a digital twin of actual traffic conditions, avoiding the high costs and safety risks of direct real-world deployment. For real-world applicability, TTC-X functions as a versatile plug-and-play system, easily adaptable to existing traffic management infrastructures, enhancing traffic monitoring and management with minimal disruption and expense.  
\end{itemize}

\section{Literature Review}

\subsection{Visual surveillance camera systems in ITS}
Intelligent transportation systems (ITS) \cite{oladimeji2023smart} have been revolutionized by various applications, including traffic surveillance \cite{aissaoui2014advanced}, data acquisition \cite{khaliq2019road,li2020trajectory}, information estimation \cite{sanguesa2015sensing,rim2011estimation}, congestion recognition \cite{han2020congestion}, and sustainably traffic management \cite{wang2021adaptive}. A diverse range of objects that can be captured through visual surveillance camera systems, including vehicles \cite{song2019vision,zhang2019vehicle}, pedestrians\cite{khalifa2020novel,el2020pedestrian}, cyclists \cite{ferreira2022identifying}, obstacles \cite{yu2020study}, road infrastructures, etc. Within the realm of traffic surveillance and monitoring, conventional methods can be found in \cite{6875912,bommes2016video,yang2018vehicle,7458203}. \cite{6875912} reviews both appearance and motion-based feature methods for vision-based vehicle detection, where the former uses the appearance features (e.g., shape, color, and texture of the vehicle) to detect the vehicle or separates it from the background and the latter one utilizes moving characteristic to distinguish vehicles from the stationary background image. They also mention popular approaches (different vehicle representations \cite{koller1993model}, Kalman filter \cite{unzueta2011adaptive,morris2008learning}, etc.) in vehicle tracking that are often used to extract vehicles’ dynamic attributes, including velocity, the direction of movement, and vehicle trajectories. \cite{bommes2016video,yang2018vehicle} give reviews on possible applications (vehicle counting, vehicle speed measurement, identification of traffic accidents, traffic flow prediction, etc.) of video-based ITS under varying challenging environments such as illumination change and different weather conditions. \cite{7458203} specifically focuses on the issues related to vehicle monitoring with cameras at road intersections. Recent innovations in computer vision (CV) and deep learning (DL) research for visual traffic surveillance systems can be found in \cite{9714212,fei2023multi}. \cite{9714212} gives a systematic review of the detection and tracking CV-based methods, while \cite{fei2023multi} conducts a comprehensive survey on multi-object multi-camera tracking (MOMCT) frameworks from multiple-object single-camera detection and tracking to vehicle re-identification approaches across cameras.

\subsection{Traffic state estimation methods}
Graph Neural Networks (GNNs) have recently gained prominence among the various methods available for traffic state estimation. By transforming large-scale road networks into graph structures, GNNs facilitate the exchange of local information between neighboring road segments. Several conventional GNN-based methods, such as Graph Convolutional Networks (GCNs) \cite{yu2020forecasting}, Graph Attention Networks (GATs) \cite{zhang2019spatial}, and Graph Recurrent Networks (GRNs) \cite{do2019effective}, have been effectively employed to forecast traffic states within large-scale road networks.
While GNN-based methods can capture the spatial dependencies of traffic states in the road network, transformer-based methods evolve as power methods of capturing the dynamic flux as traffic states are time-series data. Originating from the nature language processing (NLP) method, the transformer model \cite{vaswani2017attention} has demonstrated exceptional performance in capturing temporal dependencies within sequential data. This success has led to its adoption in addressing transportation-related challenges. Huang et al., \cite{huang2022dynamical} combined a convolutional neural network (CNN) with a modified transformer to predict network-wide traffic demand. Zhang et al.,\cite{zhang2022trajectory} proposed a spatial-temporal graph attention transformer to predict autonomous vehicle trajectories.

\subsection{Online learning and control approaches}
The proposed PiCOL methodology falls within the realm of online learning and optimization study \cite{shai_online}, also known as the multi-arm bandit problem in the literature \cite{slivkins19bandit}. The backbone of the PiCOL subscribes to exponential weights \cite{schapire99ew}, a seminal online learning algorithm, and its variant: EXP3 \cite{bianchi02adv-bandit}. Similarly to PiCOL,  \cite{Rakhlin13pred,jadbabaie15pred_opt} considers online learning with predictions of future loss functions. However, unlike our investigation into PiCOL, these prior works focus on the influence of prediction accuracy on online learning control performance without touching the interplay and mutual influence between prediction and control. In contrast to the aforementioned prior works studying single-agent cases, PiCOL is a multi-agent distributed online learning scheme, also about the line of research works on multi-agent learning \cite{tao_info, Bianchi_Lugosi_2006}. In particular, PiCOL bears a spirit of coordinating each agent's online learning through the shared loss function, similar to learning for correlated equilibria \cite{hart00regret_match,perchet14blackwell}. Finally, online learning applications in urban traffic networks remain largely uncharted. Even though a handful of prior works have applied online learning and control methods to traffic signal control \cite{hong23signal-control} and traffic forecasting \cite{guo21traffic-forecast}, our work is among the first endeavors to explore the role of online learning in PTC controls.

\section{Methodology}
\subsection{An Overview of TTC-X}\label{subsection:overview-ttcx}
\begin{figure}[t!]
    \centering
    \includegraphics[width=0.99\columnwidth]{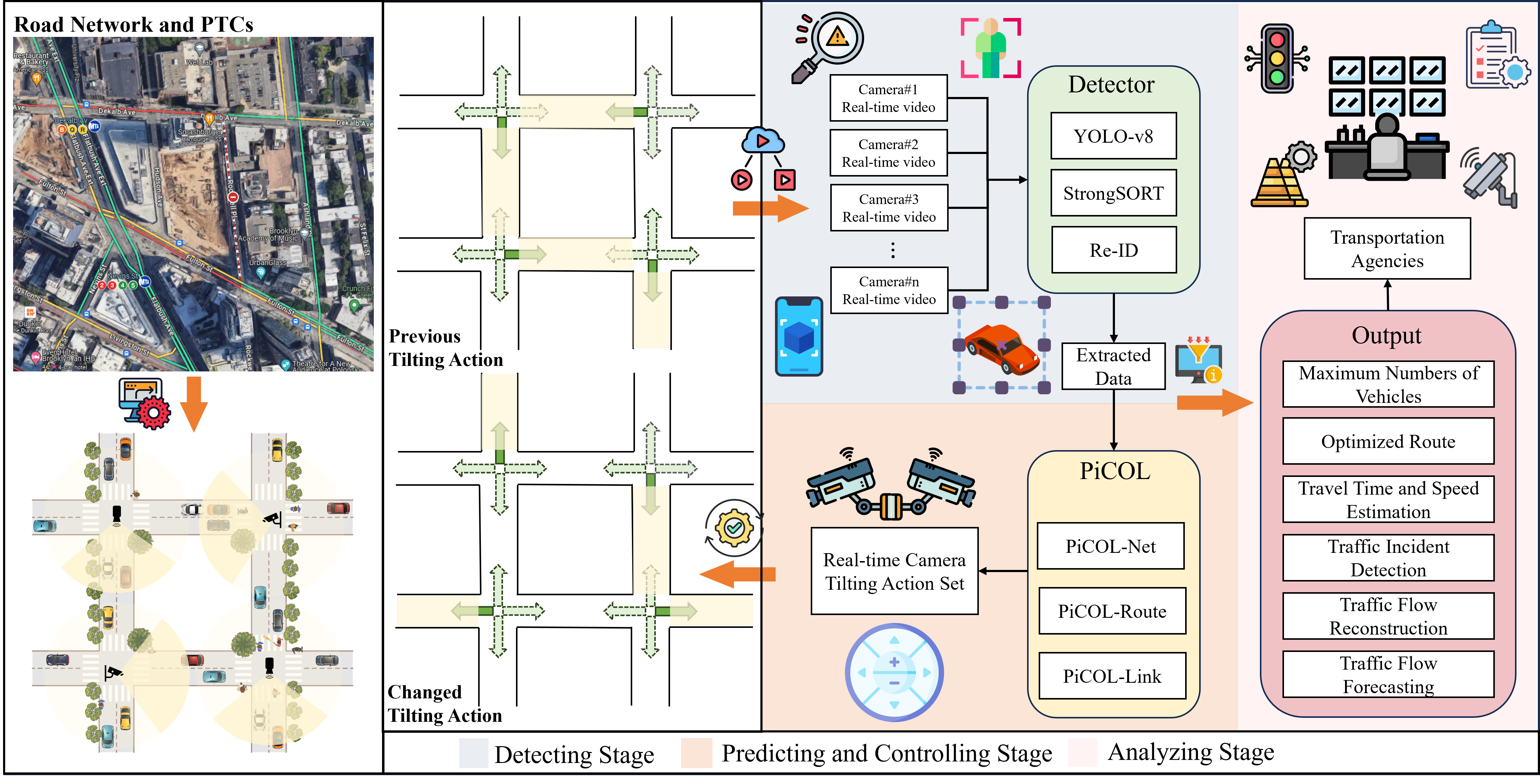}
    \caption{The overall framework architecture of TTC-X system.}
    \label{fig:overall-TTCX}
\end{figure}

The comprehensive architecture of the TTC-X system encapsulates several integral components, as shown in \Cref{fig:overall-TTCX}. The first component is the detector, which serves as the nexus for information extraction. It adeptly processes the visual information acquired through PTCs, translating into link-level (individual road segment) traffic states, such as vehicular volume, travel time, and traffic speed per road segment. The converted information is subsequently fed into the PiCOL component, which includes both the predictor and controller. Within this framework, the predictor's Spatial-Temporal Graph Prediction (STGP) module employs an offline learning paradigm, having been trained on historical data to enable real-time, online predictions. In contrast, the controller adopts an online learning approach, continuously calibrating and optimizing the PTC tilting policy in response to live traffic states. By integrating the extracted road network topology and traffic states into the STGP, the predictor acquires a spatial and temporal understanding of traffic dynamics specific to each road segment in the network.  After a warm-up phase involving PTC observation, a dynamic information feedback loop is established between the predictor's STGP and the controller, as shown in \Cref{fig:picol-overview}. The STGP provides the predictions of traffic states for several upcoming time steps. These predictions are then utilized by the controller to formulate bespoke tilting strategies, addressing diverse operational scenarios at the network, route, and link levels. Subsequent to the implementation of these strategies, the controller assimilates observations and overlays the forecasted traffic states with actual observed values. This sequence of updated traffic states is then re-introduced into the STGP for ensuing prediction cycles.  This continuous feedback mechanism allows for successive refinements in the STGP's future traffic state estimations as well as the controller's online calibration, thereby significantly augmenting the overall efficiency and precision of the TTC-X system.

\subsection{Detector}
In the proposed TTC-X system, traffic surveillance and analysis are primarily powered by deep learning models, specifically YOLOv8 for object detection and StrongSORT for object tracking. These models are fed with frames from CCTV camera data strategically positioned across the transportation network \cite{zuo2020interactive, zuo2021reference}. This section delineates the functioning of these models and their integration within the TTC-X system.

YOLOv8 (You Only Look Once, version 8) is an advanced iteration in the YOLO series\cite{bochkovskiy2020yolov4}, known for real-time object detection with high accuracy. YOLOv8 operates on the principle of dividing the input image (CCTV frames) into a grid, and each grid cell predicts bounding boxes and class probabilities. The model inputs continuing video feed from CCTV cameras and then processes each frame to detect various traffic objects (vehicles, pedestrians, cyclists). YOLOv8 is designed to identify objects swiftly and accurately, which is essential in dynamic traffic conditions. Following detection, StrongSORT \cite{du2023strongsort} is employed for tracking. It is an enhanced version of the SORT (Simple Online and Realtime Tracking)\cite{bewley2016simple} algorithm, augmented with deep learning features for improved tracking accuracy. StrongSORT receives the identified objects from YOLOv8 and tracks them across successive frames. This is crucial in monitoring traffic flow, calculating speeds, and identifying traffic incidents.
 
The integrated YOLOv8 and StrongSORT models generate multiple essential traffic data as output: vehicle counts (volume data), speeds (speed data), and behavioral patterns (safety data). This data becomes the backbone for various traffic management applications. The output is classified and channeled into three different scenarios, aligning with the multi-level surveillance strategy of TTC-X:

\begin{figure}[t!]
    \centering
    \includegraphics[width=0.5\columnwidth]{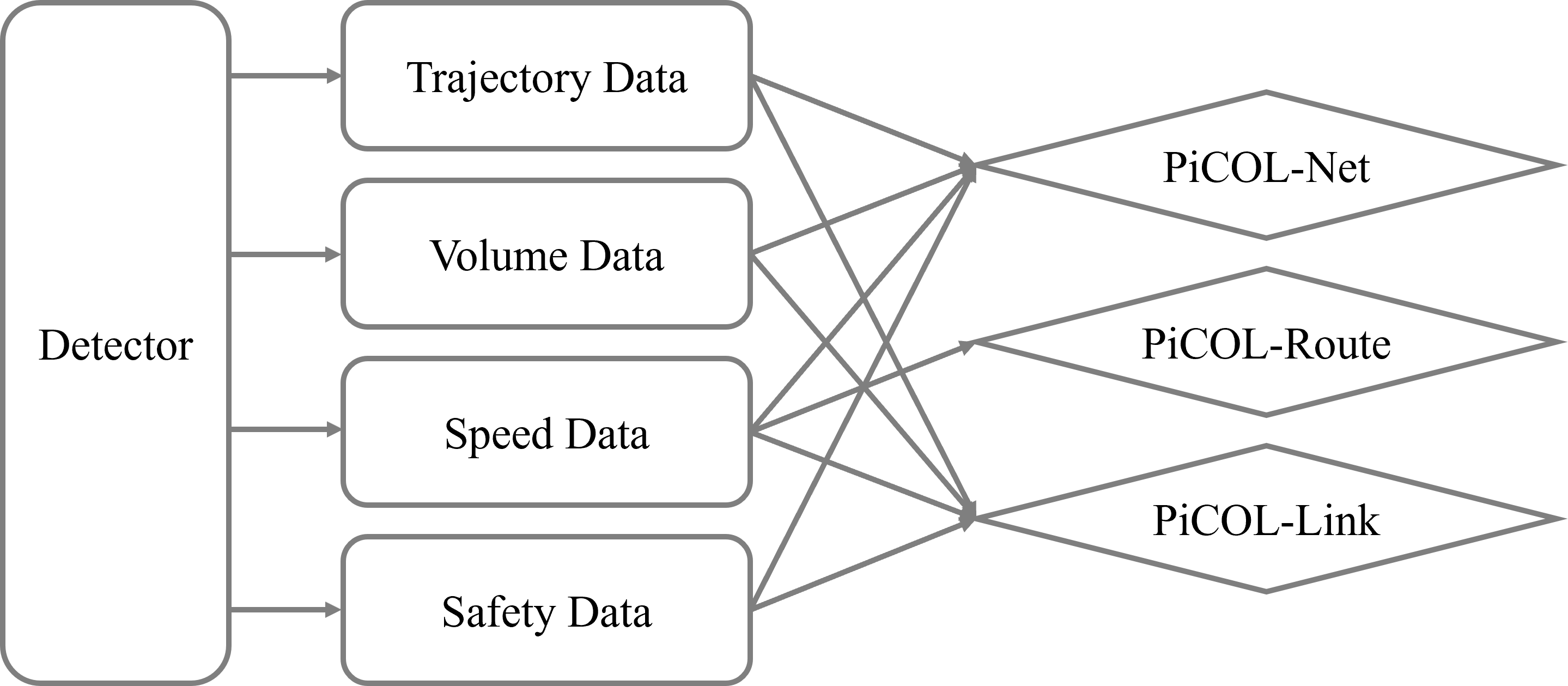}
    \caption{The data flow from the detector to the PiCOL modules.}
    \label{fig:det-flow}
\end{figure}

The Traffic-responsive Tilt Camera system (TTC-X) is engineered to optimize traffic flow, monitor routes, and forecast traffic states at different levels of urban traffic management. Figure \ref{fig:det-flow} shows the data flow from the detector to the PiCOL modules. At the network level, PiCOL-Net integrates traffic flow data to refine signal timings and ease congestion, particularly during peak periods. Route-level management via PiCOL-Route leverages this data for real-time routing, significantly improving emergency vehicle response by adapting to live traffic conditions. PiCOL-Link, at the link level, focuses on accurate estimation and forecasting of traffic states, aiding in swift incident response and maintaining traffic network efficiency.

\subsection{Predictive Correlated Online Learning}
\begin{figure}
    \centering
    \includegraphics[width=1\textwidth]{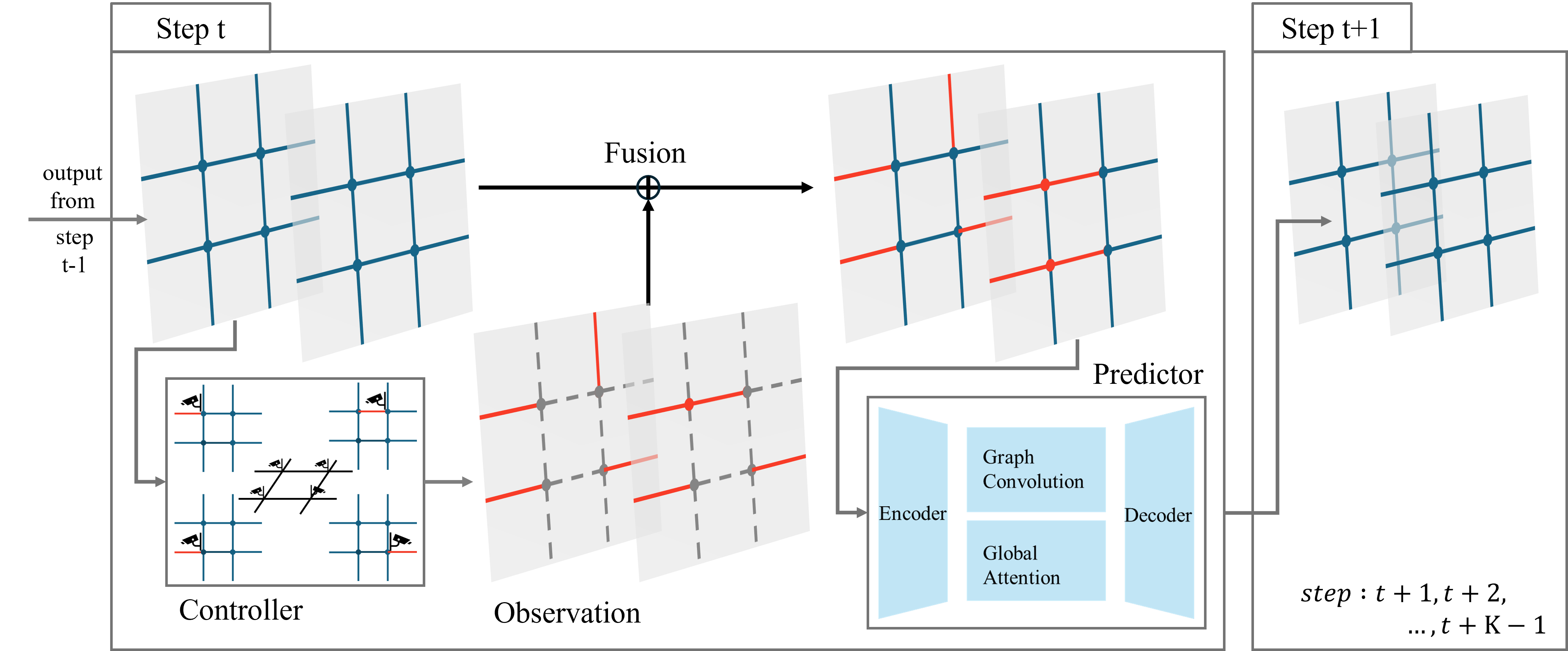}
    \caption{An overview of PiCOL module in TTC-X. The controller's tilting policy is calibrated online and distributed using the predictor's forecast of traffic states. These predictions, fused with actual traffic state observations, are fed to the predictor for the future predictions that guide the controller's calibration.  }
    \label{fig:picol-overview}
\end{figure}

The predictive correlated online learning (PiCOL) module follows the detector, which is directly responsible for the PTCs' tilting control. As shown in \Cref{fig:picol-overview}, PiCOL includes two essential pillars: the controller, the PTCs' joint online tilting policy, and the predictor, an offline-trained traffic forecasting model. Despite different data input and task objectives across the network, route, and link levels, PiCOL treats the three scenarios from a unified online learning perspective, leading to a generic formulation of these tasks as later presented in \Cref{subsubsec:multi-level} and a shared online learning control approach. The resulting online learning algorithm, detailed in \Cref{subsubsec:controller}, is a plug-and-play control scheme without any domain knowledge of the network topology and the underlying traffic evolution, readily applicable to a variety of traffic situations and road networks. Even though the online learning control is traffic and network-agnostic, it relies on the STGP to extract the spatial and temporal dependencies of traffic states (see \Cref{subsubsec:stgp}), which creates a cooperative and responsive PTC coverage of the road network.         
\subsubsection{Multi-level Real-time Traffic Information Collection and Processing}
\label{subsubsec:multi-level}
Consider a traffic network represented by a directed graph $\mathcal{G}=(\mathcal{N}, \mathcal{E})$, where the node set $\mathcal{N}:=\{1,2, \ldots, N\}$ denote the intersections, and the terms intersections and nodes are used interchangeably in this paper. Each element in the edge set $\mathcal{E}=\{(i,j)|i,j\in \mathcal{N}\}$ represents the road segment connecting two intersections. We use link, edge, and road segments interchangeably throughout the paper. We model the traffic evolution within the network during a given period of time using a discrete-time sequence $\{\bm{s}_t\}_{t\in [T]}$, where $\bm{s}_t$ denotes the real-time traffic information of interest at time $t$. The traffic state variable $\bm{s}_t$ denotes the aggregated local information from each link $s_t^{ij}$: $\bm{s}_t=[s_t^{ij}]_{(i,j)\in \mathcal{E}}$.  

PTCs are allocated to a subset of intersections for surveillance purposes. Let $\mathcal{N}^c\subset \mathcal{N}$ be the set of the nodes equipped with a PTC. With a slight abuse of notation, we denote by $i\in \mathcal{N}^c$ the PTC deployed at the $i$-th intersection. The PTC $i$ can be tilted to monitor the inbound and outbound links $(i,j)$ and $(j, i)$ from a neighboring node $j$ at a time. We define the PTC's action as a $|\mathcal{E}|$-dimensional binary vector $a^i\in \mathcal{A}^i:=\{0, 1\}^{|\mathcal{E}|}$, and the one entries correspond to the monitored edges: $a^i{(i,j)}=1$, if $(i,j)$ is covered by the PTC. Let $\bm{a}_t=\vee_{i\in \mathcal{N}^c}, a_t^i\in \{0,1\}^{|\mathcal{E}|}$ be the joint action of all cameras at time $t$, where $\vee$ denotes the entry-wise boolean operator ``or''. In plain words, $\bm{a}_t(i,j)=1$ if the edge $(i,j)$ is covered by at least one of the PTCs. Employing the entry-wise product (Hadamard product), denoted by $\otimes$, the joint observation of the TTC system is given by $\bm{s}_t^c=\bm{s}_t\otimes \bm{a}_t$, where the local traffic states without any camera monitoring are masked with zero. 
\begin{table}
    \centering
    \begin{tabular}{ll}
    \toprule
    Notation(s)  & Description\\
    \midrule
    $\mathcal{G}$, $\mathcal{N}$, $\mathcal{E}$  & The traffic network, the junction (nodes) set, and roads (edges) set  \\
    $(i,j)\in \mathcal{E}$, $i,j\in \mathcal{N}$ & The directed edge from node $i$ to $j$\\
       $T$, $[T]:=\{1,2,\dots, T\}$  & The horizon length and the set of discrete time indices  \\
       $\mathcal{N}^c$ & The subset of nodes where cameras are deployed\\
       $\mathcal{E}^{i}$ & The set of incident edges of node $i$\\
       $\mathcal{A}^{i}$, $a^i_t$ & The action set of camera $i$ and the implemented action at time $t$\\
       $\bm{s}_t=(s_t^{ij})$, $(i,j)\in \mathcal{E}$ &  The network-level traffic state information at time $t$, \\
       & an aggregation of local information on each edge\\
       $\bm{\delta}_t=(\delta_t^{ij})$, $(i,j)\in \mathcal{E}$ & The network-level relative traffic state difference at time $t$\\
       $\bm{s}_t^c $ & The joint observation of the TTC-X\\
       $\bm{\delta}_t^c$ &  The observed relative state difference\\
       $\hat{\bm{s}}_t$ & The predicted traffic state\\
       $\hat{\bm{s}}_t^f$ & The traffic state fusion, a combination of $\hat{\bm{s}}_t$ and $\bm{s}_t^c $\\
       $L(\bm{a}_t;\bm{s}_t)$ & The loss function of joint action and network state\\ 
       $\mathcal{P}_{(i_o, i_d)}$ & The set of all paths connecting the OD pair $(i_o, i_d)$\\
       $\mathcal{G}_{(i_o, i_d)}$ & The smallest subgraph containing $\mathcal{P}_{(i_o, i_d)}$\\
       $\otimes$, $\vee$ & The entry-wise product and the boolean operator ``or''\\
    \bottomrule   
    \end{tabular}
    \caption{A summary of frequently used notations.}
    \label{tab:notations}
\end{table}

Real-time traffic information can be captured and processed on multiple levels depending on the transportation agencies' needs. At the network level, the objective of the TTC-X is to monitor the edges with significant volume to collect maximum flow data at each time step. At the route level, the task is to utilize cameras to update the traffic information in real time for dynamic route recommendation. At the link level, to fulfill the traffic state reconstruction and forecasting, the TTC-X needs to be sensitive to the traffic incidents on each edge, leading to dramatic changes in traffic patterns within a short period. The cameras are tasked to capture these incidents as soon as they occur. The following formally defines the three scenarios in mathematical terms and the corresponding objective functions specifically. The key observation is that all three scenarios admit a shared optimization formulation (up to a constant). 

\paragraph{Network Level}
In the first scenario, the agency aims to capture the maximum flow in real time using the TTC-X. The traffic state ${s}_t^{ij}$ represents the traffic volume on the road segment $(i,j)$ at time $t$, while the aggregation $\bm{s}_t$ presents the network-level volume distribution. Suppose the joint action at time $t$ is $\bm{a}_t$, and then the performance metric is given by the absolute percentage error $L(\bm{a}_t;\bm{s}_t):=\|\bm{s}_t^c-\bm{s}_t\|_1/{\|\bm{s}_t\|_1}$, which evaluates the amount of traffic volume not captured by the cameras. The TTC-X aims to minimize the mean absolute percentage error (MAPE) across the horizon.
\begin{align}
\label{eq:network-loss}
    \min_{\{\bm{a}_t\}_{t\in [T]}}\frac{1}{T} \sum_{t\in [T]}L(\bm{a}_t;\bm{s}_t),\quad  L(\bm{a}_t;\bm{s}_t):=\frac{\|\bm{s}_t^c-\bm{s}_t\|_1}{\|\bm{s}_t\|_1}.
\end{align}
\paragraph{Route Level} The objective in the second scenario is real-time route planning. The cameras collectively identify the path with the least travel time in dynamic traffic patterns with unexpected traffic incidents. Such route planning admits a dynamic nature: the route recommendation shall be updated in real-time if traffic incidents render the current plan no longer optimal. In this scenario, the traffic state in information ${s}_t^{ij}$ denotes the travel time estimate of $(i,j)$, which can be obtained by the speed estimate returned by the camera.  

Consider trips traversing from $o_t$ to $d_t$ at time $t$, and the origin-destination pair $(o_t, d_t)\in \mathcal{N}\times\mathcal{N}$ admits multiple possible paths. Denote by $\mathcal{P}_{(o, d)}$ (or simply $\mathcal{P}$ if the OD pair is clear from the context)  the set of all paths connecting the $o$ and $d$, with its typical element $p\in \mathcal{P}_{(i_o, i_d)}$ represented by a sequence of nodes in which each node is connected by an edge to the next. Let $\mathcal{G}_{(o,d)}$ be the smallest subgraph containing the nodes and edges on the paths included in $\mathcal{P}_{(o,d)}$. For simplicity, we assume that those paths in $\mathcal{P}_{(o,d)}$ for any arbitrary OD pair can be fully covered by the camera nodes within $\mathcal{G}_{(o,d)}$. The travel time of the chosen path (represented by the joint action $\bm{a}_t$) is given by the inner product $ \langle \bm{a}_t, \bm{s}_t \rangle$
and hence, the TTC-X aims to return a sequence of route planning $\{\bm{a}_t\}_{t\in [T]}$ such that the total travel time is minimized, as presented in Eq.~\eqref{eq:route-loss}.
\begin{align}
\label{eq:route-loss}
    \min_{\{\bm{a}_t\}_{t\in [T]}}\sum_{t\in [T]}L(\bm{a}_t;\bm{s}_t),\quad  L(\bm{a}_t;\bm{s}_t):= \langle \bm{a}_t, \bm{s}_t \rangle,  
\end{align}
\paragraph{Link Level}
The last scenario concerns the real-time network state estimation that requires link-level change-point detection. The TTC-X is tasked to monitor the edges with significant fluctuations in traffic patterns. In this case, the quantity of interest is the relative difference in traffic volume on each edge. Let $s^{ij}_t$ be the volume number defined in the network-level case, and then the relative difference in volume at time $t$ is as $\delta^{ij}_t=(s_t^{ij}-s^{ij}_{t-1})/s^{ij}_{t-1}$. The network-level aggregation is then denoted by $\bm{\delta}_t=[\delta^{ij}_t]_{(i,j)\in\mathcal{E}}$. Similar to the network-level case,  the sensitivity of the TTC-X is evaluated through the error ${\|\bm{\delta}_t^c-\bm{\delta}_t\|_1}/{\|\bm{\delta}_t\|_1}$. Therefore, the objective function over the whole horizon is given by the time average in Eq.~\eqref{eq:link-loss}. 
\begin{align}
\label{eq:link-loss}
   \min_{\{\bm{a}_t\}_{t\in [T]}}\frac{1}{T} \sum_{t\in [T]}L(\bm{a}_t;\boldsymbol{\delta}_t),\quad L(\bm{a}_t; \bm{\delta}_t):= \frac{\|\bm{\delta}_t^c-\bm{\delta}_t\|_1}{\|\bm{\delta}_t\|_1},\bm{\delta}_t^c=\bm{\delta}_t \otimes \bm{a}_t
\end{align} 
\subsubsection{PiCOL Online-Learning Controller}
\label{subsubsec:controller}
Due to urban transportation networks' complex and highly dynamic nature, modeling the network traffic state evolution using closed-form dynamic systems is prohibitive. Even though the three distinct scenarios share a sequential decision-making formulation, the current control design methodologies, such as optimal control \cite{bertsekas2011dynamic}, rely on a well-structured dynamic model to capture the system evolution. A model-free learning-based control can better equip the cameras with responsive intelligence and online adaptability in complex and dynamic traffic patterns, enabling them to collect real-time traffic information cooperatively.         
  
With rich historical traffic data and advancement in traffic simulation and digital twins \cite{zuo2020microscopic}, reinforcement learning (RL) stands out as a strong candidate for learning intelligent TTC-X control policies. Model-free RL, such as value-based \cite{tao_multiRL} and policy-based algorithms \cite{sutton_PG}, offers a data-driven approach to tackling large-scale network control and management problems \cite{tao22confluence} without domain knowledge of underlying system dynamics. However, within the offline learning paradigm, model-free RL suffers from the distribution shift between the training and testing environments \cite{levine2020offline}. In urban transportation, the distribution shift is embodied by non-recurrent traffic incidents unobserved in the historical traffic data. An RL control policy trained using such historical data typically does not generalize well to the real-world traffic environment, which displays different traffic patterns due to incidents. Even though many research efforts, such as online meta-learning \cite{finn2019online, tao23cola},  have been dedicated to the generalization and adaptability of RL, the resulting machinery suffers from exhaustive offline training and online computation \cite{pan2023meta-sg}.

This work explores the online learning paradigm \cite{bianchi02adv-bandit, Tao_blackwell} to address the TTC-X challenges without domain knowledge of the underlying traffic and the offline training on the control policy. The notable features of the proposed PiCOL methodology, as a synthesis of an online-learning-based control algorithm (controller) and an offline-trained forecasting model (predictor), include that 
\begin{itemize}
    \item the controller employs an online learning algorithm derived from exponential weight (EW) \cite{schapire99ew}, a seminal multi-arm bandit algorithm, whose implementation is independent of the network topology and only requires the controller's online observations and predictor's forecast; 
    \item the learning-based control policy is executed and updated at each camera node in parallel, achieving distributed control;
    \item the predictor learns network-structure nature and the sequential characteristics of traffic states in offline training, and its predictions help coordinate each camera's tilting.
\end{itemize}
The following revisits the three representative scenarios and casts them as a multi-arm bandit problem, a well-established setup for online learning. Using the network-level maximum flow capturing as an example, we illustrate the operation and insights of the PiCOL.     

In the multi-arm bandit problem, initially proposed by \cite{robbins52mab}, a gambler must choose one of the finitely many slot machines to play. By pulling the arm of one of the machines, he receives a reward at each time step. Mathematically, a multi-armed bandit problem is specified by a finite action set $\mathcal{A}$ from which the decision-maker selects an action $a_t$ at each time step. After the action implementation, the environment returns a loss incurred by such action $L_t(a_t)$. Note that the loss function $L_t(\cdot)$ is time-varying and unknown to the decision-maker before time $t$. This online decision-making process can be viewed as a repeated game between the decision-maker and nature, where the game's payoff is jointly determined by the actions and the loss functions chosen by the two players, respectively, at each time step. The decision-maker aims to minimize the cumulative loss $\sum_{t\in[T]}L_t(a_t)$ by learning to play actions based on its previous online observations $\{a_1, a_2, \ldots, a_{t-1}; L_1, L_2, \ldots, L_{t-1}\}$.  

Comparing the aforementioned multi-arm bandit problem setup and the three scenarios in \Cref{subsubsec:multi-level}, we note that the latter can be viewed as a special instance of multi-arm bandit problem despite the difference in loss function definitions across the three scenarios. The joint camera action gives the finite action set: $\mathcal{A}=\prod_{i\in \mathcal{N}^c} \mathcal{A}^i$, and the time-varying loss function is specified by the traffic state: $L_t(\cdot)=L(\cdot;\bm{s}_t)$. Recasting the multi-level real-time traffic information collection and processing as a multi-arm bandit (online learning) problem brings several advantages over the control and RL approaches. {First, the multi-arm bandit formulation does not impose any statistical assumptions on the traffic state transition. As illustrated in \cite{bianchi02adv-bandit}, one can think of an oracle arbitrarily selecting the traffic state at each step without following any statistical rules, which captures the abrupt traffic disturbances. In contrast, RL formulations assume that Markovian state transition in the testing time is the same as in the training time, which can be violated if unexpected traffic incidents in the online deployment lead to traffic conditions drastically different from those in training.} {Second, as presented in the later discussion, the MAB formulation leads to an online learning control that enjoys a plug-and-play operation without offline training. Hence, such an online learning algorithm is computationally lightweight than the RL approach, which requires pre-training or fine-tuning when deployed in various traffic environments. Furthermore, the proposed online learning takes a closed-form update scheme, enabling a direct inspection of the intra-camera control coordination, which the neural network in RL cannot afford as the network parameters do not directly translate into a tangible control policy.} 

In the sequel, we unfold our PiCOL method using the network-level flow capturing as an example. The resulting learning scheme is referred to as PiCOL-Net. Our treatment below directly lends itself to the route and link-level tasks, and we refer to the corresponding learning-based control algorithms as PiCOL-Route and PiCOL-Link, respectively. If no confusion arises from the context, we simply use PiCOL for reference.  To facilitate our discussion, we begin with the assumption that the true traffic state $\bm{s}_t$ is revealed to the controller \textit{after} the joint action $\bm{a}_t$ such that the incurred loss defined in Eq.~\eqref{eq:network-loss} at each time step can be computed. This unrealistic assumption will be removed later when we use the traffic state prediction as a surrogate. Suppose the controller's joint policy at time $t$ is $\bm{\pi}_t\in \Delta(\mathcal{A})$, from which the realized joint action $\bm{a}_t$ is sampled and executed. The resulting loss is the absolute percentage error defined in Eq.~\eqref{eq:network-loss}: $L(\bm{a}_t,\bm{s}_t)=\|\bm{s}_t\otimes \bm{a}_t-\bm{s}_t\|_1/\|\bm{s}_t\|$.  Beginning with a uniform distribution over the joint action set $\bm{\pi}_1=\operatorname{Unif}(\mathcal{A})$, one can calibrate the policy using the exponential weight algorithm \cite{schapire99ew} presented in Eq.~\eqref{eq:ew} with $\gamma_t\in (0,1)$ being the discounting factor. 
\begin{align}
\label{eq:ew}
    \bm{\pi}_{t+1}(\bm{a})=\frac{\bm{\pi}_t(\bm{a}) \exp\{-\gamma_t L(\bm{a};\bm{s}_t)\}}{\sum_{\bm{a}\in \mathcal{A}}\bm{\pi}_t(\bm{a}) \exp\{-\gamma_t L(\bm{a};\bm{s}_t)\}}, \bm{a}\in \mathcal{A}.
\end{align}
EW is named after the calibration of the probability $\bm{\pi}_{t+1}(\bm{a})$ using exponential weights $\exp\{-\gamma_t L(\bm{a};\bm{s}_t)\}$. For each action $\bm{a}$, the larger the loss (error) it incurs, the smaller the exponential becomes, and consequently, the smaller its weight becomes at the next step. In plain words, the EW algorithm begins with equal probability (weights) for every possible joint camera titling action and then reduces the weights of less desired joint actions by multiplying their previous weights by the exponential.      

Simple and intuitive, the EW algorithm has proved to be an effective online learning scheme in the face of non-stationary loss sequences \cite{bianchi02adv-bandit}. The exponential weight calibration enables the TTC-X to adapt rapidly to dynamic traffic situations. However, the calibration defined in Eq.~\eqref{eq:ew} is performed centralized, which suffers from the curse of dimensionality as the number of camera nodes and associated incident edges grow. One natural remedy is to distribute the policy update and perform the calibration at each node. However, the challenge of this parallel calibration is that the joint action of all cameras determines the loss, and hence, assigning proper weights to each camera's action is the key to a distributed EW suitable for large-scale TTC-X. Inspired by game-theoretic learning methods\cite{tao22confluence}, we introduce the \textit{correlated} exponential weight (CEW) to evaluate the contribution of each individual action to the network-level loss conditional on others' actions. Mathematically, the CEW for the node $i$'s action $a^i$ at time $t$ is $\exp\{-\gamma_t L(a^i, \bm{a}_t^{-i};\bm{s}_t)\}$, given that the other nodes' joint action is $\bm{a}^{-i}_t$. Unlike the weight in centralized EW adapting the joint action to the changing traffic state for smaller loss, CEW adapts individual actions to other nodes' actions for better coordination (greater flow captured) across TTC when facing non-stationary state evolution. Define the control policy at each camera node as $\pi_t^i\in \Delta(\mathcal{A}^i)$ in Eq.~\eqref{eq:co-ew}, and the correlated exponential weight algorithm is presented in Eq.~\eqref{eq:co-ew}.
\begin{align}
\label{eq:co-ew}
    \pi_{t+1}^i(a^i) = \frac{{\pi}^i_t(a^i) \exp\{-\gamma_t L(a^i, \bm{a}_t^{-i};\bm{s}_t)\}}{\sum_{{a^i}\in \mathcal{A}^i}{\pi^i}_t(a^i) \exp\{-\gamma_t L(a^i, \bm{a}_t^{-i};\bm{s}_t)\}}, a^i\in \mathcal{A}^i.
\end{align}
In summary, the CEW algorithm above begins with equal probability for every individual action at each node and then increases the weights of actions better coordinated with other nodes' actions by multiplying their previous probability with correlated exponential weights.

Our discussion has developed so far under the assumption that the actual traffic state $\bm{s}_t$ is revealed to the controller at each time step. Yet, the controller can only receive the partial state information, $\bm{s}_t^c=\bm{s}_t\otimes \bm{a}_t$, captured by the cameras. Consequently, the exponential weights of actions not employed cannot be computed, i.e., $\exp\{-\gamma_t L(a^i, \bm{a}_t^{-i};\bm{s}_t)\}$, for $a^i\neq a^i_t$, are unknown, which prevents CEW algorithm in Eq.~\eqref{eq:co-ew} from real-world deployment. 

The case of missing data in online observation necessitates the usage of network-wide traffic estimation. Taking inspiration from spatial-temporal forecasting methods \cite{zhang2024semantic}, we develop a traffic state forecasting model, referred to as the predictor, by combining graph convolution network \cite{guo2021learning} and transformer \cite{vaswani2017attention}. \Cref{subsubsec:stgp} presents a detailed description, and we here abstract from this machine learning machinery and treat it as a mapping $\mathscr{F}$ to complete our algorithm development on PiCOL. 

During the online implementation, the predictor $\mathscr{F}$ takes in a batch of past partial observation $\{\bm{s}_\tau^c\}_{\tau=t-K}^{t-1}$ and predicts a batch of future traffic states of the same batch size, denoted by $\{\hat{\bm{s}}_\tau\}_{\tau=t}^{t+K-1}$. In other words, the predictor forecasts future K-step traffic state evolution based on the most recent K-step observations. At the $t$-th step, we first carry out a data fusion operation that integrates the actual partial observation $\bm{s}_t^c$ and the prediction $\hat{\bm{s}}_t$. This fusion operation fills the missing data point in $\bm{s}^c_t$ with the corresponding entries from $\hat{\bm{s}}_t$, meaning that the prediction $\hat{s}_t^{ij}$ serves as the surrogate for the missing local state information $s_t^{ij}$, for $(i,j)$ not covered by any cameras. Recall that the joint action $\bm{a}_t$ is a binary vector of $|\mathcal{E}|$-dimension, and its non-zero entries indicate the links covered by at least one camera at time $t$. Let $\mathbbm{1}$ be the $|\mathcal{E}|$-dimensional all-one vector, the fusion state, denoted by $\hat{\bm{s}}_t^f$, is then define as $\hat{\bm{s}}_t^f=\bm{s}_t^c+\hat{\bm{s}}_t\otimes(\mathbbm{1}-\bm{a}_t)=\bm{s}_t\otimes \bm{a}_t+ \hat{\bm{s}}_t\otimes(\mathbbm{1}-\bm{a}_t)$, which is a substitute of the true state. Consequently, one can compute the CEW using the fusion, leading to the proposed PiCOL scheme: 
\begin{subequations}
\begin{align}
    \hat{\bm{s}}_t  &= \mathscr{F}(\{\hat{\bm{s}}^f_\tau\}_{\tau=t}^{t+K-1}),\\
    \hat{\bm{s}}_t^f  &=\bm{s}_t^c+\hat{\bm{s}}_t\otimes(\mathbbm{1}-\bm{a}_t),\\
    \pi_{t+1}^i(a^i) &=(1-\epsilon_t) \frac{{\pi}^i_t(a^i) \exp\{-\gamma_t L(a^i, \bm{a}_t^{-i};\hat{\bm{s}}^f_t)\}}{\sum_{{a^i}\in \mathcal{A}^i}{\pi}^i_t(a^i) \exp\{-\gamma_t L(a^i, \bm{a}_t^{-i};\hat{\bm{s}}^f_t)\}} + \epsilon_t \frac{1}{|\mathcal{A}^i|}, a^i\in \mathcal{A}^i.\label{eq:exp3}
\end{align}
\end{subequations}
One may notice that the PiCOL scheme admits an additional $\epsilon$-exploration term in Eq.~\eqref{eq:exp3}, subscribing to the seminal online learning algorithm \textit{exploration and exploitation with exponential weights} (EXP3) \cite{bianchi02adv-bandit}. As we later present in \Cref{subsec:interplay}, the exploration term is instrumental in PiCOL-Link, encouraging the PTC to probe every edge with positive probabilities. As such, TTC-X is more likely to directly capture the sudden change in link-level traffic conditions or infer the incidents by observing fluctuations on neighboring edges through the predictor. In general, the explorations contribute to accurate predictions, which, in turn, benefits online policy calibration. To fully explain the interconnection between the controller and the predictor, we first articulate the inner workings of the predictor in the subsequent.  

\subsubsection{Spatial-Temporal Graph Predictor (STGP)}\label{subsubsec:stgp}

The motivation for introducing the STGP into PiCOL is similar to the data imputation task in traffic forecasting problems. Traffic data imputation, which is often used to deal with the failure of traffic sensing systems, spatial-temporal forecasting methods \cite{zhang2024semantic} are preferred due to their effectiveness in learning the network-structure nature as well as the sequential characteristics of traffic states. Similarly, in the context of PiCOL, deploying a limited number of PTCs within an extensive transportation network often results in underrepresentation. This limitation arises because each PTC can monitor only one direction at a time, necessitating confident estimations for unobserved directions.

\paragraph{Traffic State Estimation Problem Formulation} The formulation of STGP is straightforward, adhering to the notations in Table.\ref{tab:notations}. The traffic states of road segments on the traffic network $\mathcal{G}$ at time step $t$ are denoted by traffic state matrix $s_t = (s_t^{ij})$, where $(i,j)\in \mathcal{E}, i,j \in \mathcal{N}$. Given a sequence of recent historical traffic states ${\{\bm{s}_\tau\}_{\tau=t-K}^{t-1}}$ over the past $K$ time steps, STGP aims to predict the sequence of future traffic states, denoted by $\{\hat{\bm{s}}_\tau\}_{\tau=t}^{t+K-1}$,  over the next $K$ time steps. It is important to note that STGP is predicting traffic states $\{\hat{\bm{s}}_\tau\}_{\tau=t}^{t+K-1}$ over each road segment, which is each edge in $\mathcal{G}$, denoted by $(i,j)\in \mathcal{E}$.

The design of STGP is open-ended and flexible; one requirement of STGP is to capture the spatial dependencies among road segments, acknowledging that traffic propagates in short- and long-range within the road network. Another favor of STGP is to capture both stationary and non-stationary temporal trends in the evolution of traffic states. This study takes the encoder-decoder architecture following the design of vanilla transformer \cite{vaswani2017attention} and integrates the graph convolution kernel and global attention kernel inspired by \cite{guo2021learning}. 

\begin{figure}[t!]
    \centering
    \includegraphics[width=0.9\columnwidth]{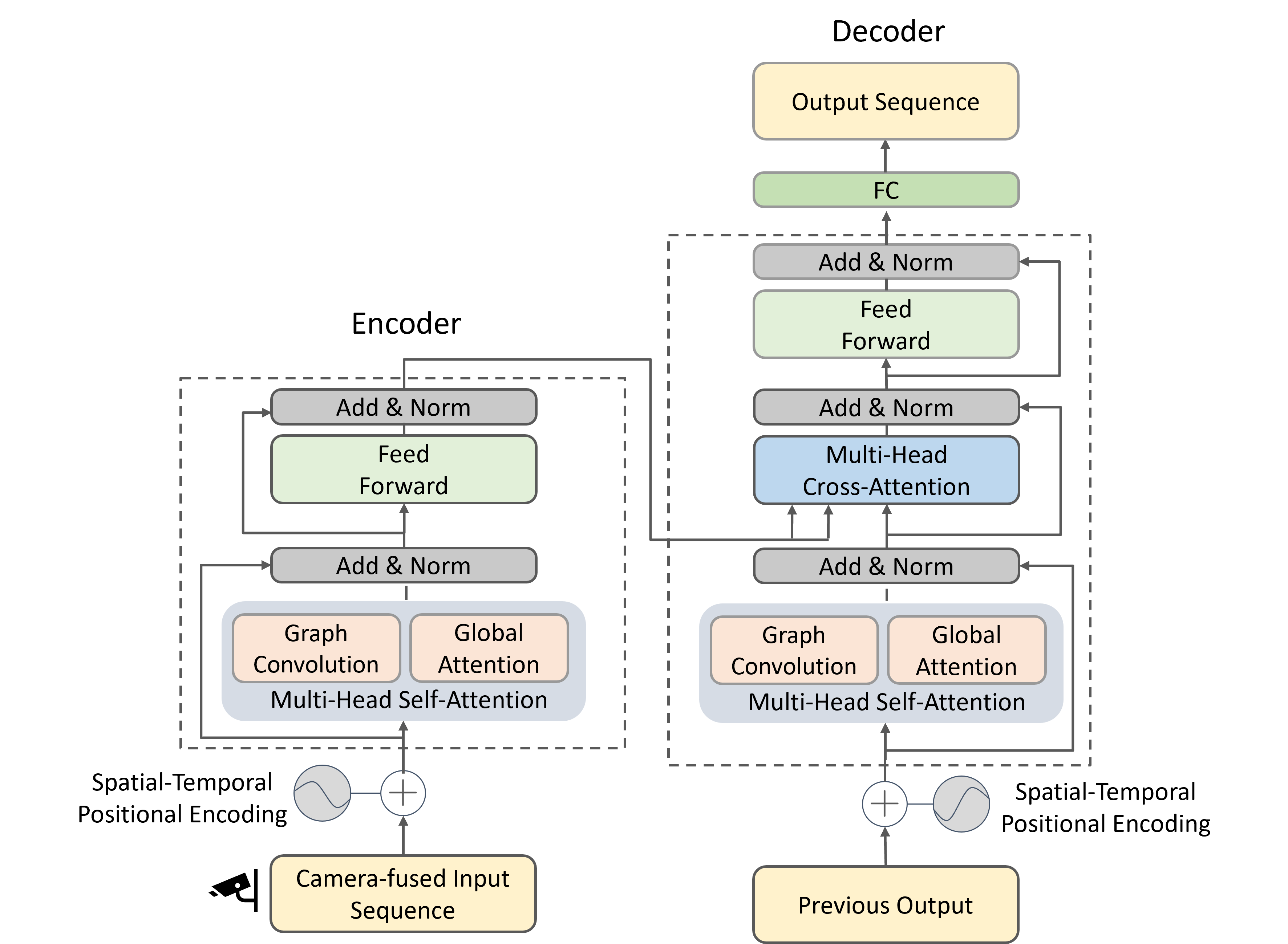}
    \caption{The architecture of STGP in TTC-X.}
    \label{fig:STGP-arch}
\end{figure}

\paragraph{Encoder-decoder architecture} The encoder transforms the input sequence $\{{\bm{s}_\tau}_{\tau=t-K}^{t-1}\}$ into a latent representation $\{\Tilde{\bm{s}}_\tau\}_{\tau=t-K}^{t-1}$, which the decoder then uses to predict $\{\hat{\bm{s}}_\tau\}_{\tau=t}^{t+K-1}$ with masked attention. This process employs a moving window of length $K$ in an auto-regressive manner, where each time step's prediction builds upon the last. As shown in \Cref{fig:STGP-arch}, both the encoder and decoder comprise a multi-head self-attention mechanism, with the decoder incorporating an additional cross-attention sublayer to engage with the encoder's output.

\paragraph{Graph convolution kernel} The original design of the graph convolution kernel follows the well-established layer-wise propagation rule proposed in \cite{kipf2016semi}, which can be formally defined as follows for directed graph:

\begin{equation}
\label{Eq: GCN}
    \begin{aligned}
       H^{(l+1)} = \sigma \big( \Tilde{D} ^{-1} \Tilde{A} H^{(l)} W^{(l)} \big).
    \end{aligned}
\end{equation}
\noindent where $H^{(l)} \in \mathbb{R}^{E \times d_{\text{model}}}$ is the graph signal representation in the $l^{th}$ layer. It is important to note that in this study, matrix $A$ describes the connectivity between edges $(i,j)\in \mathcal{E}$, rather than between nodes, as the node set $\mathcal{N}$ represents each intersection. $E$ is the number of edges in $\mathcal{E}$. $\Tilde{A} = A + I_E$ is the adjacency matrix of the directed graph $G$ with added self-connections. $I_E \in \mathbb{R}^{E \times E}$ is the identity matrix. $D$ is degree matrix and $W^{(l)}$ is layer-specific trainable weights. $\sigma(\cdot)$ is the nonlinear activation function. 

Given the dynamic nature of real-world transportation networks, this study adapts the spatial correlation weight matrix to learn spatial dependencies dynamically \cite{guo2021learning}. The spatial weight matrix $\bm{S}$ is thus defined to modify $A$ through an element-wise dot-product operation:

\begin{equation}
\label{Eq: SA matrix}
    \begin{aligned}
       \mathbf{S} = \text{softmax} \Big(\frac{H^{(l)}H^{(l)\top}}{\sqrt{d_{\text{model}}}}\Big) \in \mathbb{R}^{E \times E}.
    \end{aligned}
\end{equation}

\begin{equation}
\label{Eq: SA matrix}
    \begin{aligned}
       \text{GCN}(H^{(l)}, \Tilde{A}) = \sigma \Big( \big((\Tilde{D} ^{-1} \Tilde{A})\odot \mathbf{S} \big) H^{(l)} W^{(l)} \Big).
    \end{aligned}
\end{equation}
 
\paragraph{Global attention kernel} Although the design of STGP is open-ended, it is not arbitrary. Given the constraints in PiCOL with a limited number of cameras as well as partial observations, it is favorable for STGP to estimate traffic states of unobserved road segments based on observed information from other road segments. While graph convolution captures the local (i.e., short-range) spatial dependencies residing in the traffic propagation, the long-range patterns also exist in traffic states along the road network. The multi-head attention mechanism allows each position in the traffic state sequence to attend to every other position, enabling the transformer to represent long-range patterns in traffic states \cite{vaswani2017attention}. Furthermore, acknowledging the complexity of traffic patterns often subject to non-stationary disturbances, the STGP needs to extend beyond the standard multi-head attention. It incorporates 2D convolutional layers instead of the linear projection in the original multi-head attention mechanism. This change aims to extract the local trend of temporal patterns \cite{guo2021learning}, as shown in Eq.\ref{Eq: conv2d multi-head}.

\begin{equation}
\label{Eq: conv2d multi-head}
    \begin{aligned}
       \text{MultiHead}(\mathbf{Q}, \mathbf{K}, \mathbf{V}) & = \oplus \big(\text{head}_1, \cdots, \text{head}_h\big) W^O,\\
      \text{head}_i & = \text{Attention}\big(\Phi ^Q_i* \mathbf{Q}, \Phi ^K_i * \mathbf{K}, \mathbf{V}W^V_i\big),
    \end{aligned}
\end{equation}
\noindent where $\Phi ^Q_i$ and $\Phi ^K_i$ are the parameters for the 2D convolutional layers, the query $Q \in \mathbb{R} ^{L_Q \times d_{\text{model}}}$, key $K\in \mathbb{R} ^{L_K \times d_{\text{model}}}$ and value $V \in \mathbb{R} ^{L_V\times d_{\text{model}}}$ are both transformed from the input token sequence $X \in \mathbb{R} ^{L \times C}$, and $L_Q, L_K, L_V$ are the corresponding length. Linear projection $W_i^V \in \mathbb{R}^{d_{\text{model}} \times d_k}$ maps dimension of $d_{\text{model}} $ into head $h_i$.




\subsection{The Interplay between Controller and Predictor}
\label{subsec:interplay}
The spatial and temporal dependencies captured by STGP are the cornerstone of a successful PiCOL implementation, contributing to network-wide camera coordination and rapid response. Recall that the online calibration relies on a proper evaluation of CEW in Eq.~\eqref{eq:exp3}, which further depends on accurate predictions of traffic states on unobserved road segments. Thanks to the graph convolution in STGP, the predictor can infer one edge's missing local traffic state with high accuracy using observations from its neighboring edges. In other words, the fusion state $\hat{\bm{s}}_t^f$  gives a decent substitute to the true state $\bm{s}_t$ and, consequently, effective exponential weights in online learning. 

In addition to the spatial aspect, STGP also learns offline the temporal correlation among the time series of traffic states through the global attention mechanism. The combination of multi-head attention with 2D convolutional layers enables STGP to capture local temporal trends and long-range (global) temporal dependencies among traffic states. When some cameras observe unanticipated traffic incidents, the learned local temporal trend helps STGP accurately forecast the future network traffic evolution within a short window under these abnormal conditions. Receiving the corresponding predictions, PiCOL quickly adjusts its policy calibration and responds rapidly to these unexpected traffic conditions by tilting neighboring cameras to those affected edges. One important note is that certain exploration in PiCOL (see Eq.~\eqref{eq:exp3}) is necessary when dealing with unexpected traffic conditions. Incidents may likely take place on lesser-monitored edges with relatively small flow. Hence, PiCOL needs to assign non-zero weights to every edge for link-level change-point detection. In contrast to the local trend, the long-range temporal dependencies in STGP unveil the long-term evolution pattern caused by recurrent traffic conditions (e.g., rush hours) to the controller. Therefore, PiCOL prepares for these recurrent incidents by tilting cameras in relevant directions without worrying about losing track of important information from unobserved edges.

\section{Results and discussion}

\subsection{Dataset summary}
\begin{figure}[t!]
    \centering
    \includegraphics[width=0.9\columnwidth]{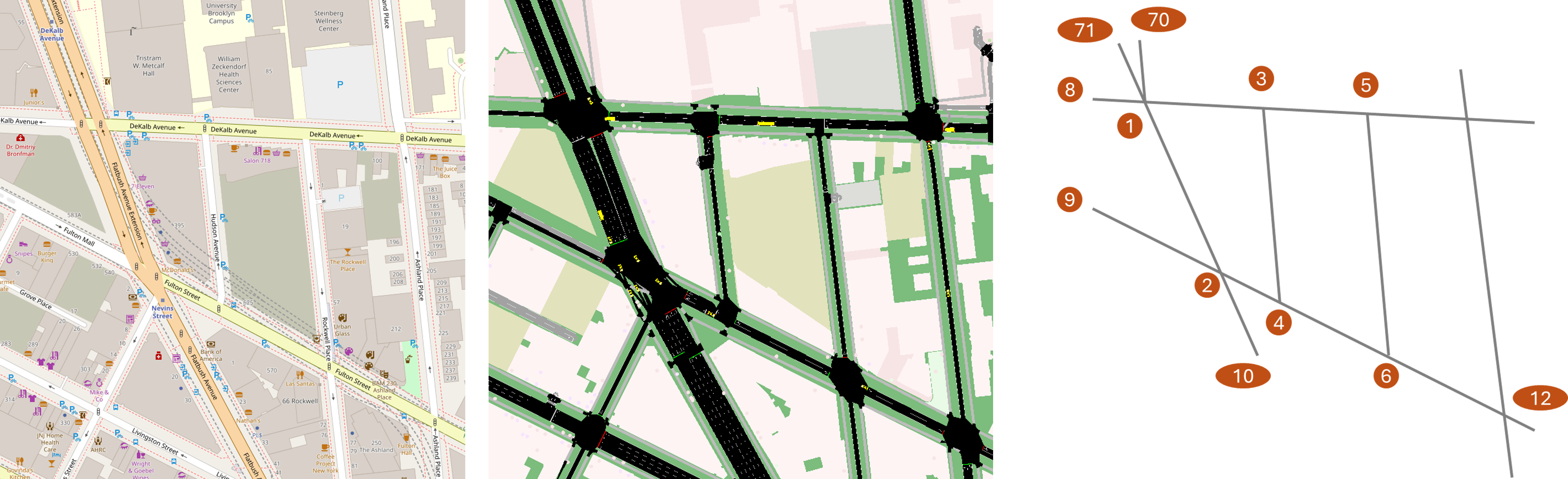}
    \caption{The simulation network building process. Left to right: Open Street Map, Network in SUMO, Abstracted Network.}
    \label{fig:map-gen}
\end{figure}

\paragraph{Simulation Environment} The dataset critical to the training and experimental phases of our TTC-X is generated using the Simulation of Urban Mobility (SUMO) \cite{SUMO2018}. SUMO is a versatile, open-source traffic simulation package designed to handle large networks and provide detailed representations of vehicular movements. This approach models each vehicle individually, allowing us to capture intricate interactions and behaviors in real-world traffic. This level of detail is essential for understanding and predicting traffic dynamics, especially in response to disruptions or changes in typical traffic patterns. The simulated environment for our study is meticulously crafted to mirror the dynamics of the Flatbush Avenue corridor, spanning from Dekalb Avenue to Fulton Street. This corridor is a microcosm of urban traffic complexity, making it an ideal candidate for our simulation needs. The process of network generation, depicted in Figure \ref{fig:map-gen}, involves a meticulous recreation of this urban corridor within the SUMO environment.

\paragraph{Calibration of the Simulation} The realism and accuracy of the simulation are paramount. The simulation is calibrated using real-world data from multiple sources to achieve this. This data encompasses various aspects of urban traffic, such as travel times, volumes, speeds, and environmental factors like weather conditions. Additionally, surrogate safety measurements are included to provide a more comprehensive understanding of traffic dynamics and safety scenarios. This multi-faceted calibration ensures that our simulation closely mirrors actual traffic conditions and behaviors observed on Flatbush Avenue.

\paragraph{Dataset Generation} For the foundation of our dataset, we generated data representing 200 days of base traffic conditions. Each day in the simulation is detailed, encompassing 24 hours with a granularity of one-second step length. This high-resolution data provides an intricate view of daily traffic patterns and fluctuations.

We introduced change point data into the simulation to simulate traffic disruptions and study their impact. This involved randomly closing lanes for durations varying from one to five hours, thus creating 38 unique sets of change point scenarios. These disruptions are crucial for testing the robustness and responsiveness of the TTC-X  under varying traffic conditions. The output from these simulations is comprehensive. For each edge in the network, we generated data that includes average volume and average speed, recorded on a minute-by-minute basis. This granularity allows for a nuanced understanding of traffic flow and speed patterns across different segments of the road network.

\begin{figure}[t!]
    \centering
    \includegraphics[width=0.95\columnwidth]{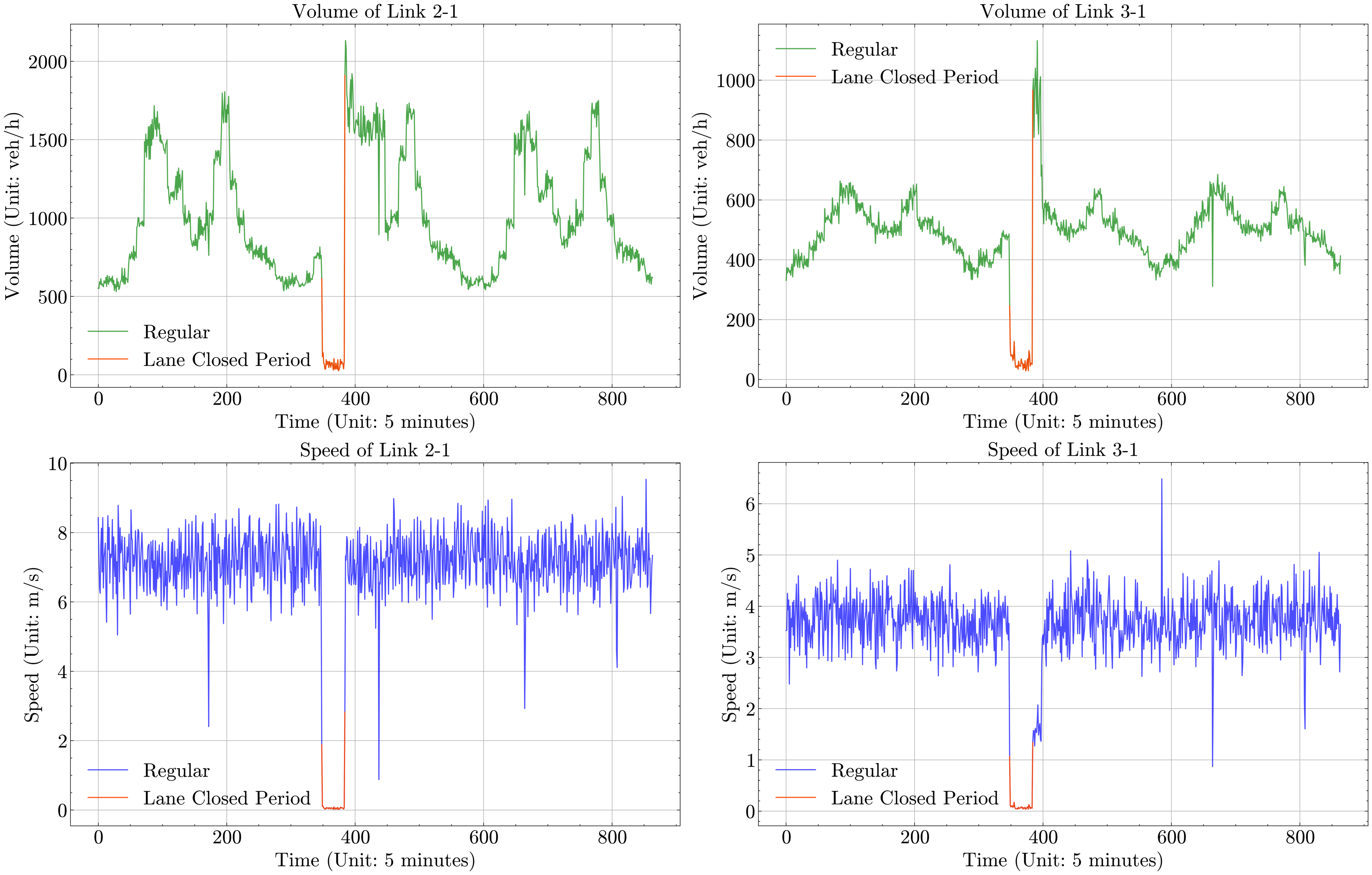}
    \caption{The volume and speed data sample for different links.}
    \label{fig:data-vis}
\end{figure}

\subsection{Experiment Setup}\label{subsec:exp-setup}
\paragraph{Predictor Configuration} The predictor in the TTC-X, based on the Spatial-Temporal Graph Prediction (STGP) model (Section.\ref{subsection:overview-ttcx}), follows an offline learning approach. Data is divided into training, validation, and test sets in a 6:2:2 ratio and normalized to the range $[-1,1]$ using min-max normalization. STGP implementation uses the Pytorch library \cite{paszke2019pytorch} on a Windows-11 desktop with an AMD Ryzen 9 7900X, 64.0 GB RAM, and an NVidia RTX 3090Ti. Optimal hyperparameters determined from validation tests include eight attention heads ($h=8$), a learning rate of $0.001$, and a model dimension of $d_{\text{model}}=64$. Traffic state data is aggregated into 5-minute intervals, with both input and prediction sequences spanning one hour (12-time steps), balancing detail and computational efficiency for effective traffic state prediction.

\paragraph{Controller Configuration} 
In accordance with the minute-by-minute data generation, the controller also operates on a minute-by-minute basis in the simulation network, and the horizon length is $T=1440$ (a day). Since each prediction instance denotes the average traffic state forecast for a 5-minute interval, the controller uses the one prediction instance repeatedly five times within the interval until a new prediction is generated for the next 5-minute interval. The motivation behind this 5-minute predictor operation cycle is that calling the predictor usually consumes 0.8-1 seconds. Hence, we choose a 5-minute cycle instead of a minute-by-minute cycle to lower the time complexity in experiments. As mentioned above, the predictor's batch size is $K=12$, so the input sequence spans one hour. Consequently, the first hour of the day is the warm-up period, where each PTC uniformly picks one edge to monitor at each time step. The starting time is 1 am, and the proposed PiCOL algorithm takes over afterward.   

For the network-level maximum flow capture task, we set the exploration rate of PiCOL-Net to zero, whereas the rate becomes 0.3 in the route-level and link-level tasks. In particular, we compare the PiCOL-Link's performance under 0-exploration and 0.3 exploration in the link-level forecasting and reconstruction tasks to showcase the role of exploration. On a separate note, we consider purely predictor-based forecasting and reconstruction, in addition to PiCOL-Link with 0/0.3-exploration. After the warm-up, the predictor-based method discards the online observations and only uses past predictions to forecast and reconstruct the link-level traffic state. All PiCOL variants under different tasks share the same discounting factor 1. Unless otherwise specified, all results on MAE and MAPE are obtained from 20 repeated experiments with different traffic simulation data and random seeds.  All experiments are carried out on the same desktop mentioned above. 
\subsection{Network-level: Maximum Flow Capture}
To evaluate PiCOL-Net's capability in network-level flow capturing, we consider the maximum number of vehicles captured by PTCs in TTC-X. Towards this end, we employ two metrics: mean absolute error (MAE) and mean absolute percentage error (MAPE) of the observed volume over the true volume within the $k$-th hour. Using notations defined in \Cref{tab:notations}, the two metrics are defined in Eq.\ref{eq:obs-maemape}, where $k$ stands for the hour index. 
\begin{equation}
\label{eq:obs-maemape}
    \begin{aligned}
    &ObsMAE_k= {1}/{60} \sum_{t=60 k}^{60(k+1)-1}{\|\bm{s}^c_t-\bm{s}_t\|_1} \\
    &ObsMAPE_k={1}/{60} \sum_{t=60 k}^{60(k+1)-1}{\|\bm{s}^c_t-\bm{s}_t\|_1}/{\|\bm{s}_t\|_1}
\end{aligned}
\end{equation}

\begin{equation}
\label{eq:fu-maemape}
    \begin{aligned}
    &FusionMAE_k= {1}/{60} \sum_{t=60 k}^{60(k+1)-1}{\|\hat{\bm{s}}^f_t-\bm{s}_t\|_1} \\
    &FusionMAPE_k={1}/{60} \sum_{t=60 k}^{60(k+1)-1}{\|\hat{\bm{s}}^f_t-\bm{s}_t\|_1}/{\|\bm{s}_t\|_1}
\end{aligned}
\end{equation}

In addition to observation data, we also consider the maximum traffic flow estimated by the TTC-X, reflecting the total travel demand within the road network, which can also be split into inbound and outbound travel demand. With the knowledge of network-wide travel demand, traffic managers can better understand the current traffic situations and make corresponding operation or planning strategies. Same as the argument above, we use MAE and MAPE of the fusion data over the true data as evaluation metrics, and the definitions are shown in Eq.\ref{eq:fu-maemape}, respectively. The lower the MAPE is, the less information is lost regarding estimated traffic demand within the road network. 

\begin{figure}[ht]
    \centering
    \includegraphics[width=1\textwidth]{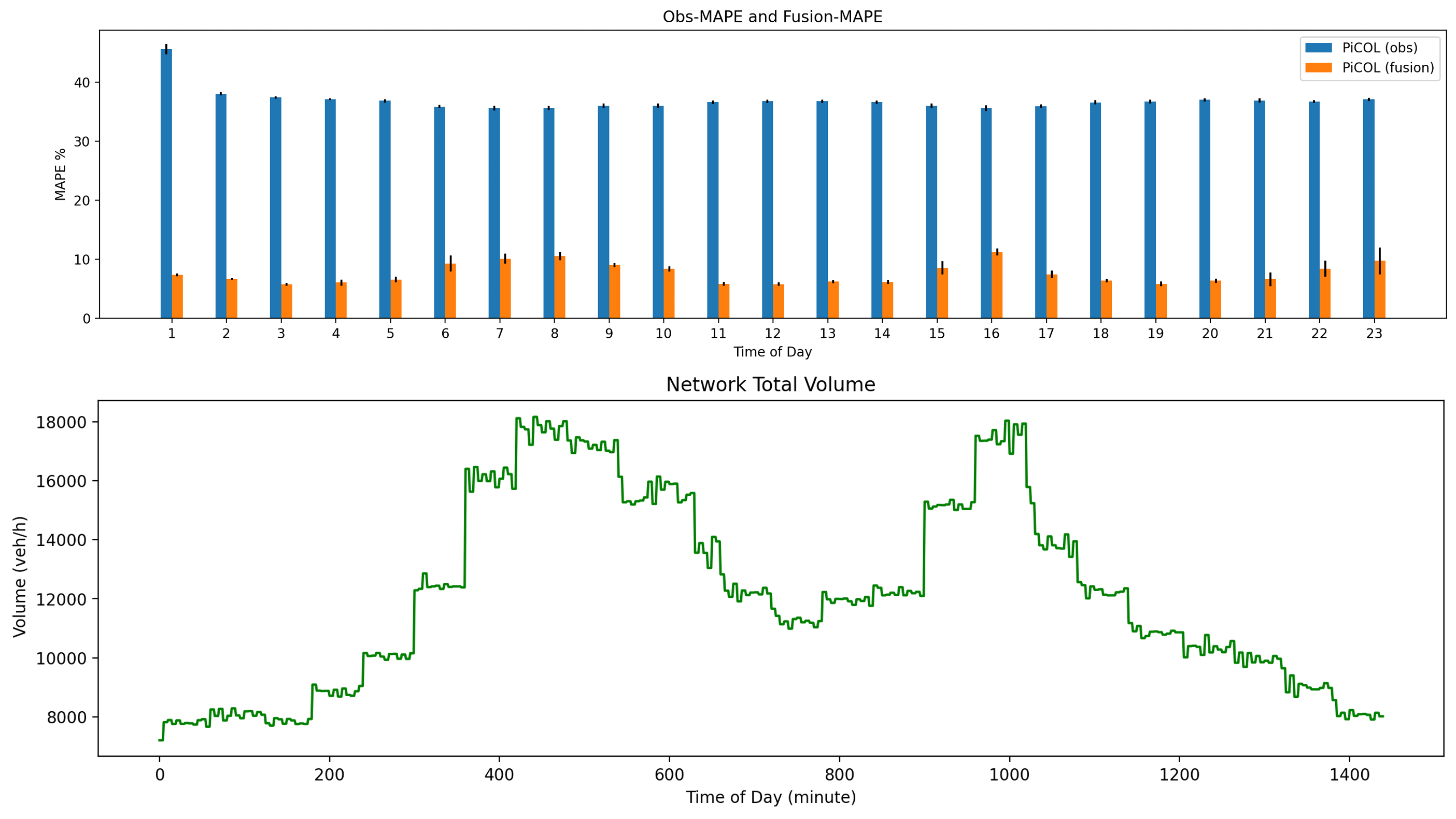}
    \caption{The bar plot of the hourly MAPE for OBS and Fusion. Employing PiCOL-Net, TTC-X can capture over 60\% of the total number of vehicles every time of day when facing dynamic traffic patterns. Moreover, the fusion data offer highly accurate estimates of the underlying traffic volumes.    }
    \label{fig:mape-bar-net}
\end{figure}

\Cref{tab:mape-mae-netlevel} summarizes the evaluation of the four metrics. The lower part of \Cref{fig:mape-bar-net} indicates the distribution of network-wide total traffic volume throughout the day; it is found that the total network-wide traffic volume ranges between 6000 vehicle/hour and 20000 vehicle/hour with clear temporal patterns. By a close look at $ObsMAPE$, one can see that TTC-X can capture over 60\%  of the total number of vehicles within the road network every time of day, as visualized in \Cref{fig:mape-bar-net}. Despite the variant of temporal patterns in traffic volume evolution shown in \Cref{fig:mape-bar-net}, TTC-X, powered by PiCOL, always keeps track of road segments with a significant number of vehicles, maintaining $ObsMAPE$ and $ObsMAE$ around a stable level. This stable performance is credited to the PTCs' coordination achieved by the correlated policy calibration in PiCOL, where PTCs cooperatively monitor the road network and adjust their policies to the dynamic traffic patterns in coordination. Furthermore, PTCs' online observations of those significant road segments under PiCOL policies serve as informative inputs to the predictor, which gives accurate inferences on unobserved edges with relatively small flows using the spatial and temporal dependencies. The inferences, in turn, help the controller to update the tilting strategies online. The interplay between the predictor and the controller enables TTC-X to derive an accurate estimation of total travel demand for every time of day, as evidenced by the low $FusionMAPE$ across the whole day (less than 10\%) in \Cref{tab:mape-mae-netlevel}.
The results presented in this section emphasize the cost-efficiency of the TTC-X. Using just six PTCs, TTC-X successfully detected over 60\% of the total vehicle count within the road network. Moreover, it estimated the network-wide traffic volume with a Mean Absolute Percentage Error (MAPE) of less than 10\%. This demonstrates the effectiveness of TTC-X in providing comprehensive traffic monitoring and estimation with a minimal camera setup.
\begin{table}[ht]
\small
    \centering
    \begin{tabular}{rlccccc}
    \toprule
       &  & 1 am & 2 am  & 3 am & 4 am & 5 am  \\
      \midrule
       \multirow{2}{*}{\rotatebox[origin=c]{90}{Obs}}&  MAPE & $45.60\pm 0.87 $ & $38.01\pm 0.27$ & $37.40\pm 0.19$  & $37.11\pm 0.15$  & $36.83\pm 0.28$   \\
         & MAE  & $3700.29 \pm 68.57$  & $2985.74 \pm 20.99$ & $3315.45 \pm 16.98$  & $3739.35 \pm 15.01$ & $4577.32 \pm 35.17$ \\
         \midrule
       \multirow{2}{*}{\rotatebox[origin=c]{90}{\footnotesize Fusion}} & MAPE  & $7.38 \pm 0.22$  & $6.66 \pm 0.15$ & $5.77 \pm 0.25$ & $6.08 \pm 0.50$ & $6.59 \pm 0.51$ \\
      & MAE  & $596.92 \pm 17.40$  & $521.24 \pm 11.61$&  $512.31 \pm 22.07$ & $614.03 \pm 50.01$ & $821.63 \pm 66.57$  \\
       \bottomrule
      & &6am & 7 am & 8 am & 9 am & 10 am \\
        \midrule
       \multirow{2}{*}{\rotatebox[origin=c]{90}{Obs}}& MAPE & $35.86\pm0.28$ &$35.58\pm 0.40$   &  $35.62\pm 0.38$    & $ 36.00\pm 0.38$   & $36.00\pm 0.37$  \\
       & MAE &$5777.06\pm 45.28$  & $6334.23 \pm 72.11$  & $6129.33 \pm 65.31$& $5608.93 \pm 58.16$  & $5274.64 \pm 54.10$  \\
       \midrule
      \multirow{2}{*}{\rotatebox[origin=c]{90}{\footnotesize Fusion}} & MAPE& $9.28\pm 1.37$  & $10.12 \pm 0.84$  & $10.57 \pm 0.73$    & $9.03 \pm 0.38$  & $8.40 \pm 0.44$ \\
      & MAE& $1512.86\pm 220.45$  & $1819.67 \pm 150.10$  & $1834.54 \pm 125.20$& $1425.07 \pm 59.34$  & $1247.56 \pm 66.60$   \\
       \bottomrule
      &  & 11 pm & 12 pm & 13 pm & 14 pm & 15 pm \\
       \midrule
      \multirow{2}{*}{\rotatebox[origin=c]{90}{Obs}}& MAPE  &$36.59\pm 0.28$ & $36.79\pm 0.28$ &$36.75\pm 0.26$ & $36.62\pm 0.28$& $35.98\pm 0.40$\\
       & MAE  &$4485.29\pm 33.91$ & $4140.98\pm 31.13$ &$4397.28\pm 30.92$ &$4477.97\pm 33.75$ & $5452.79\pm 60.79$\\
       \hline
     \multirow{2}{*}{\rotatebox[origin=c]{90}{Fusion}}& MAPE  &$5.90\pm 0.29$ & $5.80\pm 0.29$ & $6.23\pm 0.27$ & $6.18\pm 0.35$ & $8.58\pm 1.12$\\
       & MAE  &$727.78\pm 40.41$ & $653.53\pm 34.55$ & $748.26\pm 33.83$ &$756.16\pm 42.32$ & $1310.91\pm 170.30$\\
      \bottomrule
      &  & 16 pm & 17 pm & 18 pm & 19 pm & 20 pm \\
        \midrule
      \multirow{2}{*}{\rotatebox[origin=c]{90}{Obs}}& MAPE  &$35.60\pm 0.46$ & $35.91\pm 0.34$ & $36.57\pm 0.36$ & $36.68\pm 0.32$ & $37.02\pm 0.27$\\
       & MAE  & $6238.18\pm 80.32$ & $5076.98\pm 48.38$ & $4488.97\pm 43.12$ &$3993.01\pm 35.14$ & $3834.64\pm 28.10$\\
       \midrule
     \multirow{2}{*}{\rotatebox[origin=c]{90}{Fusion}}& MAPE  & $11.28\pm 0.60$& $7.47\pm 0.66$ & $6.40\pm 0.27$ & $5.86\pm 0.43$ & $6.40\pm 0.37$\\
       & MAE  & $1987.12\pm 104.12$ & $1068.34\pm 93.14$ & $788.01\pm 34.88$ & $636.69\pm 45.42$ & $656.43\pm 36.00$\\
	\bottomrule
       & & 21 pm & 22 pm & 23 pm & & \\
       \midrule
      \multirow{2}{*}{\rotatebox[origin=c]{90}{Obs}}& MAPE  &$36.90\pm 0.33$ & $36.72\pm 0.26$ & $37.07\pm 0.25$& &\\
       & MAE  &$3688.67\pm 33.27$ & $3327.74\pm 23.33$ & $2003.48\pm 20.73$ & &\\
       \midrule
     \multirow{2}{*}{\rotatebox[origin=c]{90}{Fusion}}& MAPE  & $6.65\pm 1.15$& $8.43\pm 1.38$& $9.75\pm 2.26$ & &\\
       & MAE  & $659.71\pm 111.69$ & $753.52\pm 119.97$ & $779.91\pm 181.60$ & &\\
        \bottomrule
    \end{tabular}
    \caption{The MAPE and MAE of Obs and fusion data per hour. Despite the time-varying traffic pattern within a day, PiCOL-Net always maintains a stable ObsMAPE around 40\% and FusionMAPE of less than 10\%. Given that the total network-wide traffic volumes range from 6000 veh/h and 20000 veh/h, the ObsMAE and FusionMAE indicate that PiCOL-Net successfully captures a significant share of total vehicle numbers for every time of day.}
    \label{tab:mape-mae-netlevel}
\end{table}

\subsection{Route-level: Dynamic Route Planning}
\begin{figure}[t!]
    \centering
    \includegraphics[width=0.95\columnwidth]{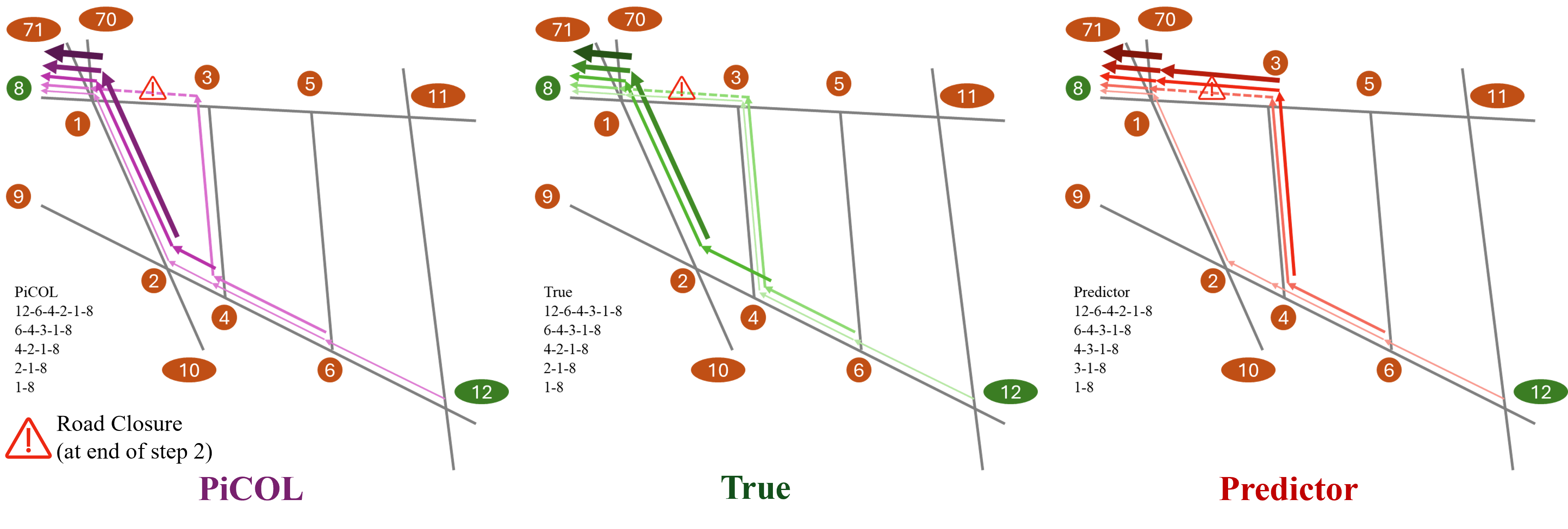}
    \caption{Dynamic route planning using PiCOL, Predictor, and True data. In this case, real-time dynamic routes are planned for trips traversing from origin Node 12 to destination Node 8, shown in green color.}
    \label{fig:optimalroute}
\end{figure}

To evaluate the performance of TTC-X at the route level, we examine PiCOL-Route's capability of dynamic route planning in a special testing case where an unexpected traffic incident takes place within the route planning time window. Specifically, we consider trips that need to traverse from origin Node 12 to destination Node 8 in the road network (\Cref{fig:optimalroute}) starting from the $290^{th}$ min, and an entire lane closure event takes place 10 min later. When vehicles arrive at a new intersection (a new node), we update the route recommendation with a new route using the predictor (i.e., derive the optimal path with the least travel time using prediction data), the proposed PiCOL (using the fusion data with online observations included), and the actual travel time information. We report the recommended route sequence and associated estimated time of arrival (ETA) provided by the three approaches above in \Cref{tab:route-time}, and a visualization of the route sequence is presented in \Cref{fig:optimalroute}.

In \Cref{fig:optimalroute}, each route recommendation is indicated by colored arrows, with darker shades representing more recent recommendations. For example, the lightest purple arrows in the PiCOL plot indicate the route recommended to the vehicle when arriving at Node 12. Comparing the routes and time estimates under the three methods, we notice that PiCOL's initial planning (12-6-4-2-1-8) is the same as the predictor's, which differs from the true optimal path. This is because PiCOL largely relies on the prediction to figure out each path's travel time when there is insufficient information during the initial exploration. As PiCOL begins to calibrate its policy using online observations and explore different paths, the time estimate gets closer to the truth (see the $2^{nd}$ row in \Cref{tab:route-time}). Thanks to PiCOL's rapid response to real-time traffic fluctuations, it timely captures the lane closure on $3-1$ and swiftly replans the route that bypasses the closed road segment when the vehicle arrives at node 4. In contrast, the predictor fails to respond to the incident and continues to recommend 4-3-1-8, and the estimate deviates much further from the truth.         

This case study underscores TTC-X's adaptability and real-time response ability in real-time dynamic route planning, which is crucial for managing unexpected traffic situations like major lane closure events. PiCOL's online policy calibration allows TTC-X to swiftly adjust to such disruptions, providing real-time optimal routing with accurate travel time estimates. This responsiveness helps vehicles avoid traffic incidents effectively. In comparison, GPS-based navigation methods rely heavily on the density of probe vehicles to aggregate travel time for each road segment. This navigation is slow in response to sudden changes compared to TTC-X, which feeds real-time high-resolution information to PiCOL and makes fast and accurate decisions on route planning. This capability is not only beneficial for routine traffic management but is also critical in life-saving scenarios involving emergency vehicles. In such situations, the high accuracy and real-time operation provided by TTC-X are essential for efficient traffic management, ensuring that emergency responders can navigate through traffic with minimal delay.  

\begin{table}[ht]
\small

\renewcommand{\arraystretch}{1.2}
\begin{tabular}{llllllllll}
\toprule
                      &                   & \multicolumn{2}{c}{Predictor} &  & \multicolumn{2}{c}{PiCOL} &  & \multicolumn{2}{c}{Truth} \\ \cline{3-4} \cline{6-7} \cline{9-10}
                      &                   & Route             & ETA      &  & Route          & ETA     &  & Route           & ETA    \\
                      \midrule
\multirow{5}{*}{\rotatebox[origin=c]{90}{ $\longleftarrow$time}} & \multirow{2}{*}{\rotatebox[origin=c]{90}{Before}} &\multicolumn{1}{:l}{ 12-6-4-2-1-8}      & 25 min 34.44 s     &  & 12-6-4-2-1-8   & 25 min 29.33 s    &  & 12-6-4-3-1-8    & 25 min 62.14 s   \\
                      &                   &\multicolumn{1}{:l}{  6-4-3-1-8 }        & 20 min 28.56 s    &  & 6-4-3-1-8      & 20 min 40.68 s    &  & 6-4-3-1-8       & 20 min 58.31 s   \\
                      & \multirow{3}{*}{\rotatebox[origin=c]{90}{After}} & \multicolumn{1}{|l}{ 4-3-1-8 }          & 15 min 20.98 s     &  & 4-2-1-8        & 15 min 100.95 s   &  & 4-2-1-8         & 15 min 99.29 s   \\
                      &                   &\multicolumn{1}{|l}{ 3-1-8 }            & 10min 9.80 s      &  & 2-1-8          & 10min  23.63 s   &  & 2-1-8           & 10min  23.63 s   \\
                      &                   &\multicolumn{1}{|l}{ 1-8 }              & 5 min      &  & 1-8            & 5 min     &  & 1-8             & 5 min   \\
                      \bottomrule
\end{tabular}
\renewcommand{\arraystretch}{1}
\caption{Sequence of route recommendations and corresponding ETAs under Predictor, PiCOL, and the true data. As time proceeds, PiCOL-Route acquires enough online observations, leading to more accurate ETAs starting from Node 6. When the lane closure suddenly happens on 3-1, PiCOL-Route is able to rapidly respond to this incident by re-planning the route, as shown in the $3^{rd}$ row.  }
\label{tab:route-time}
\end{table}

\subsection{Link-level: Traffic State Forecasting and Reconstruction}

\begin{figure}
    \centering
    \includegraphics[width=1\textwidth]{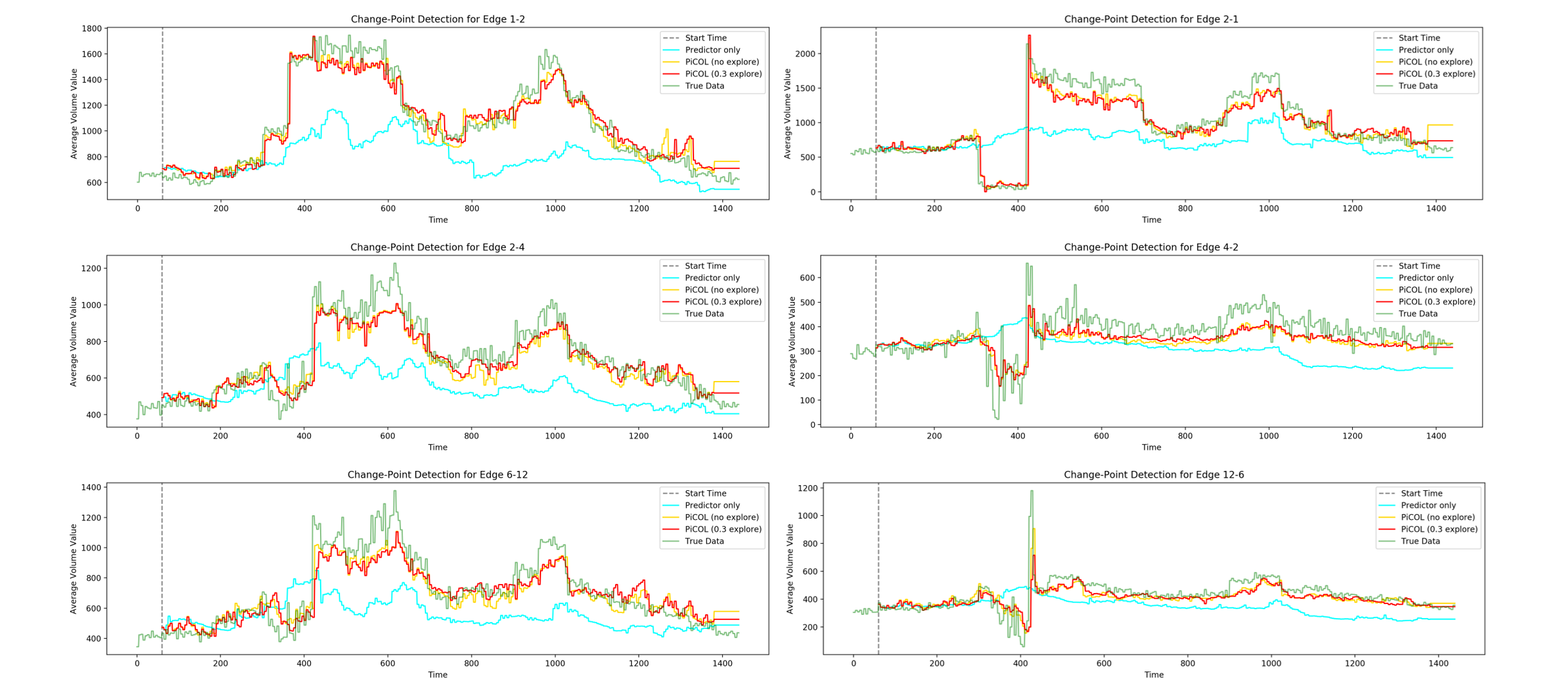}
    \caption{Selected edge results for link-level forecasting task. PiCOL-3 outperforms PiCOL-0 in general, particularly regarding the forecast on Edge 2-1 and its neighboring Edge 2-4 and 4-2. }
    \label{fig:link-forecast}
\end{figure}

In this section, we assess the direct output of PiCOL-Link, which estimates traffic flow for individual road segments within the network. As outlined in \Cref{subsubsec:multi-level}, PiCOL-Link's loss function emphasizes the relative differences in traffic volume in two consecutive time steps, particularly focusing on segments with significant traffic fluctuations (e.g., abrupt increases or decreases in volume).
At the link level, we define two key sub-tasks: 1) the forecasting task and 2) the reconstruction task. The forecasting task leverages the predictive capabilities of the predictor to forecast future traffic volumes, utilizing input data refined by the controller's observations. This task empowers traffic managers with foresight, allowing them to anticipate potential congestion or identify segments prone to traffic build-up, thus enabling early response and resource allocation to mitigate congestion. The reconstruction task, on the other hand, aims to depict real-time traffic evolution across each road segment. Its goal is to reconstruct traffic flow to match the actual state closely, demonstrating TTC-X's capacity to utilize a limited number of PTCs for comprehensive network coverage, underscoring its cost efficiency.
We evaluated PiCOL-Link using a one-day traffic dataset, including a full-lane closure on Edge 2-1 starting at 300 minutes and lasting 2 hours. As illustrated in the top right of \Cref{fig:link-forecast}, this incident led to a sharp traffic volume decrease at 300 minutes, with a rebound around 420 minutes post-event. \Cref{tab:link-mae} displays the edge-wise Mean Absolute Error (MAE), defined in Eq.~\ref{eq:fore-re-mae}, for both forecasting and reconstruction tasks across various road segments, revealing accurate results with MAEs ranging from 12.36 to 371.96 vehicles/hour for forecasting and 1.25 to 344.54 vehicles/hour for reconstruction. 
\begin{equation}
\label{eq:fore-re-mae}
    \begin{aligned}
        &ForecastingMAE^{ij} =1/1440 \sum_{t=0}^{1439}|\hat{s}_t^{ij}-{s}_t^{ij}|,\\
        &ReconstructionMAE^{ij} = 1/1440 \sum_{t=0}^{1439}|{\hat{s}_t}^{f,ij}-{s}_t^{ij}|, \text{ for} (i,j)\in \mathcal{E}.
    \end{aligned}
\end{equation}

Comparing two PiCOL settings (refer to \Cref{subsec:exp-setup}), PiCOL-3 generally outperforms PiCOL-0 in both tasks, particularly at Edge 2-1 and neighboring edges affected by the lane closure. This outcome highlights PiCOL-Link's design strategy of incorporating exploration to enhance coverage and responsiveness to sudden disturbances.

\begin{figure}
    \centering
    \includegraphics[width=1\textwidth]{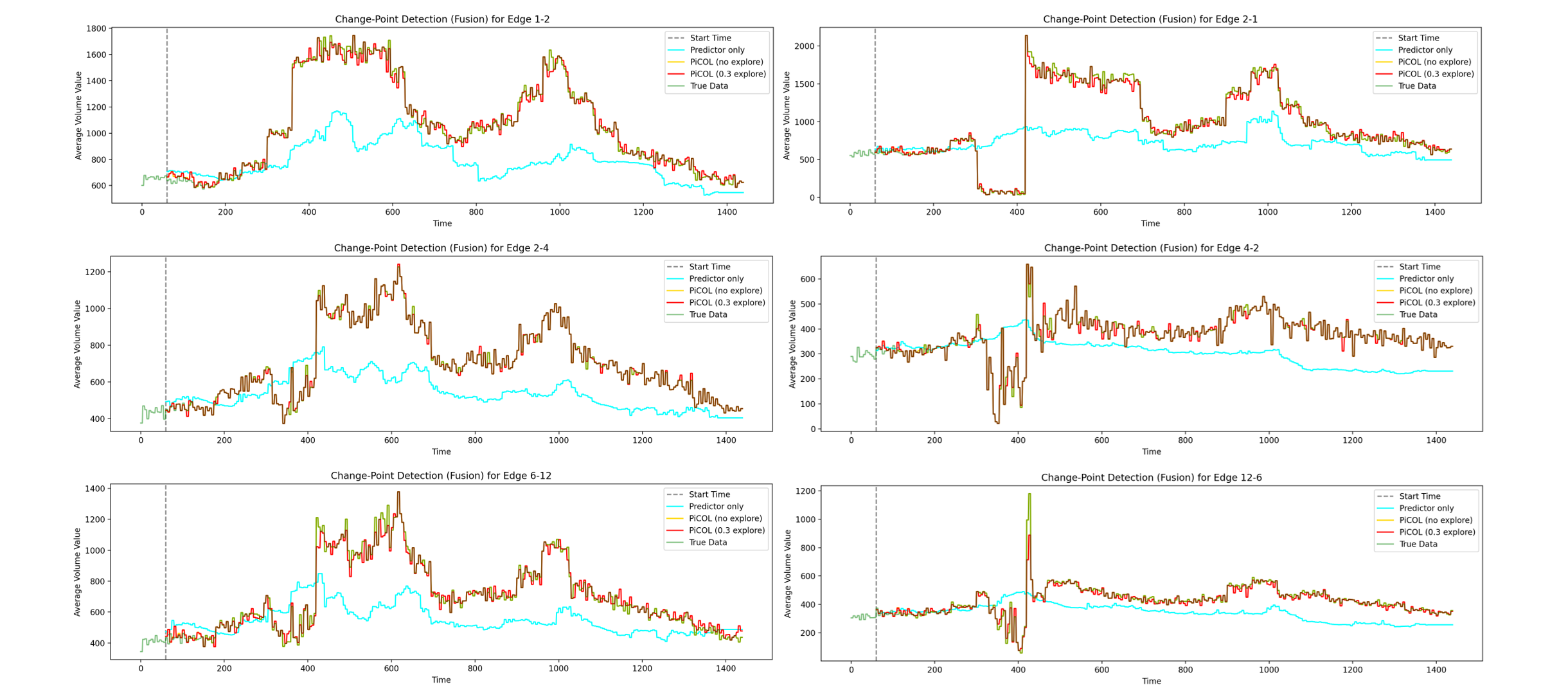}
    \caption{Selected edge results for link-level reconstruction task. Both PiCOL instances achieve impressive reconstruction accuracy. PiCOL-3 slightly outperforms on Edge 2-1 and its neighboring edges. }
    \label{fig:link-reconstruct}
\end{figure}

Visualizations in \Cref{fig:link-forecast} and \Cref{fig:link-reconstruct} display selected edge results for both tasks. {The green line represents the ground truth traffic volume. The orange line (PiCOL-0) and the red line (PiCOL-3 with 0.3 exploration) show the predictive results from STGP using real-time fusion data as input. In contrast, the blue line illustrates the outcomes from STGP when it does not use real-time fusion information, relying solely on the initial input during the warm-up period and proceeding with self-rotating movement.} The vertical dashed line indicates the start time index for the PiCOL-Link module after the warm-up period from 0 AM to 1 AM, as detailed in \Cref{subsec:exp-setup}. {It is evident in \Cref{fig:link-forecast} and \Cref{fig:link-reconstruct} that the orange and red lines, which incorporate real-time fusion data, significantly outperform the blue line. This demonstrates that the real-time fusion of traffic observations from the controller can greatly enhance the link-level traffic state estimation compared to relying solely on historical data.}

\begin{figure}
    \centering
    \includegraphics[width=0.99\columnwidth]{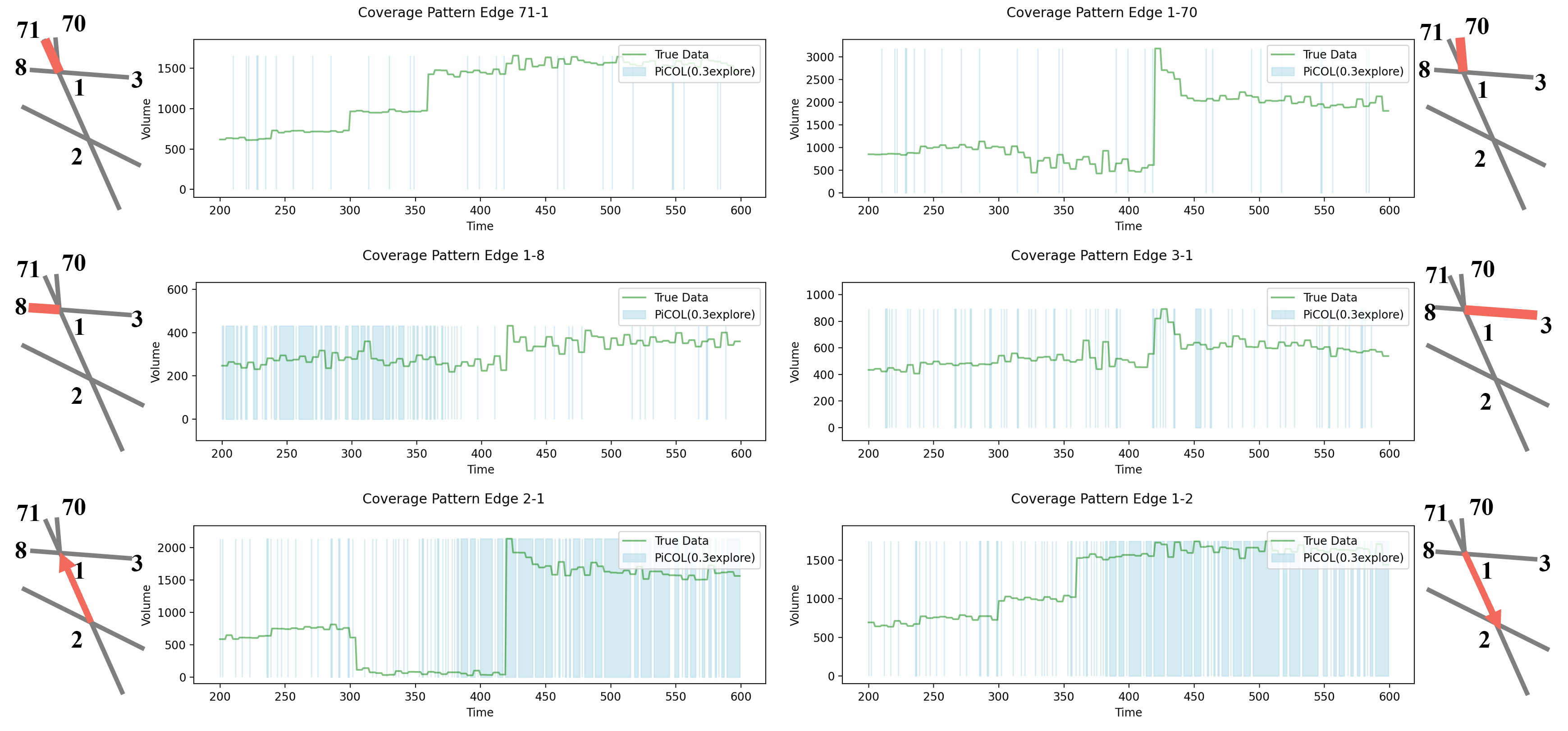}
    \caption{PTC coverage patterns at Intersection 1 (Node 1) under PiCOL-3. The light blue bars indicate the PTC coverage at each time step, showing that PiCOL successfully captured the lane closure on Edge 2-1 and the subsequent queue buildup on its upstream Edge 4-2. }
    \label{fig:link-action_plot}
\end{figure}

PiCOL accurately captured the lane closure event on Edge 2-1 and the subsequent queue buildup on its upstream Edge 4-2, demonstrating PiCOL-Link's rapid response capability within TTC-X to sudden traffic changes. This is further corroborated in \Cref{fig:link-action_plot}, which shows PTC coverage changes at Node 1. The green line depicts real traffic volume on each edge, and the light blue vertical bars indicate PTC coverage at Intersection 1, which is capable of monitoring six edges. This finding confirms TTC-X's ability to detect and quickly respond to traffic disturbances, with its focus on monitoring Edge 2-1 around the time of the lane closure event.

\begin{table}[ht]
\small
    \centering
    \begin{tabular}{rlccccc}
    \toprule
       &  & 1-2 & 1-70  & 1-8 & 2-1 & 2-4    \\
      \midrule
       \multirow{2}{*}{{Forecasting}}
         & PiCOL-0  & $206.43 \pm 75.45$  & $414.19 \pm 69.96$ & $74.22 \pm 5.00$  & $287.19 \pm 90.57$ & $\mathbf{125.61 \pm 7.86}$  \\
      & PiCOL-3  & $\mathbf{156.31 \pm 26.39}$  & $\mathbf{371.96 \pm 18.14}$ &  $\mathbf{73.78 \pm 2.21}$ & $\mathbf{223.14 \pm 33.31}$ & $126.75 \pm 4.29$   \\
      \hline
      \multirow{2}{*}{{Reconstruction}} & PiCOL-0  & $146.32 \pm 131.95$  & $393.43 \pm 125.17$ & $\mathbf{45.28 \pm 34.62}$  & $195.47 \pm 180.14$ & $\mathbf{17.64 \pm 31.29}$ \\
      & PiCOL-3  & $\mathbf{75.87 \pm 50.60}$ & $\mathbf{344.54 \pm 17.64}$& $53.55 \pm 21.19$ & $\mathbf{99.72 \pm 71.73}$ & $30.71 \pm 21.15$   \\
       \bottomrule
      & & 2-10 & 3-1 & 4-2 & 4-3 & 4-6 \\
        \midrule
       \multirow{2}{*}{{Forecasting}}
       & PiCOL-0 & $294.23 \pm 77.81$  & $101.93 \pm 14.97$& $\mathbf{107.22 \pm 2.34}$  & $22.56 \pm 1.62$ & $\mathbf{134.92 \pm 19.86}$   \\
      & PiCOL-3 & $\mathbf{264.95 \pm 38.72}$  & $\mathbf{86.86 \pm 4.95}$& $107.84 \pm 1.91$  & $\mathbf{22.47 \pm 1.08}$ & $135.90 \pm 12.99$ \\
      \hline
      \multirow{2}{*}{{Reconstruction}} & PiCOL-0  & $186.50 \pm 173.74$  & $100.57 \pm 14.99$ &  $\mathbf{12.39 \pm 22.03}$ & $\mathbf{2.78 \pm 4.56}$ & $\mathbf{48.09 \pm 66.87}$ \\
      & PiCOL-3  & $\mathbf{177.82 \pm 103.09}$ & $\mathbf{72.40 \pm 4.79}$ & $24.53 \pm 16.70$ & $5.47 \pm 0.42$ & $62.68 \pm 44.74$   \\
       \bottomrule
      &  & 5-3 & 5-6 & 6-4 & 6-12 & 71-1 \\
       \midrule
      \multirow{2}{*}{Forecasting}
       & PiCOL-0  & $86.23 \pm 15.66$  & $\mathbf{12.36 \pm 1.44}$& $\mathbf{86.59 \pm 8.51}$  & $154.00 \pm 27.34$ & $256.22 \pm 66.21$ \\
      & PiCOL-3 & $\mathbf{70.21 \pm 4.82}$  & $12.37 \pm 1.43$& $89.11 \pm 8.98$  & $\mathbf{148.82 \pm 12.16}$ & $\mathbf{208.92 \pm 17.79}$  \\
      \hline
      \multirow{2}{*}{{Reconstruction}} & PiCOL-0  & $80.80 \pm 24.13$  & $\mathbf{1.25 \pm 0.15}$ & $\mathbf{30.15 \pm 38.50}$  & $\mathbf{82.26 \pm 82.85}$ & $248.06 \pm 85.57$ \\
      & PiCOL-3  & $\mathbf{57.33 \pm 4.49}$ & $3.16 \pm 0.30$ &  $42.40 \pm 29.14$ & $192.16 \pm 52.29$ & $\mathbf{194.29 \pm 16.67}$   \\
      \bottomrule
      &  & 9-2 & 10-2 & 11-5 & 12-6 &   \\
        \midrule
      \multirow{2}{*}{{Forecasting}}
       & PiCOL-0  & $105.35 \pm 8.48$  & $287.25 \pm 85.46$& $102.13 \pm 17.36$  & $83.07 \pm 11.23$ &     \\
       & PiCOL-3  & $\mathbf{102.01 \pm 4.36}$  & $\mathbf{272.61 \pm 56.19}$& $\mathbf{84.82 \pm 6.34}$  & $\mathbf{82.19 \pm 7.69}$ &  \\
       \hline
      \multirow{2}{*}{{Reconstruction}} & PiCOL-0  &  $101.39 \pm 20.88$ & $\mathbf{182.21 \pm 178.35}$ & $101.60 \pm 17.45$  & $\mathbf{44.11 \pm 41.27}$ &  \\
      & PiCOL-3  & $\mathbf{94.88 \pm 4.07}$ & $185.79 \pm 118.02$& $\mathbf{76.94\pm 5.62}$ & $52.13 \pm 28.57$ &    \\
       \bottomrule
    \end{tabular}
    \caption{The MAE of forecasting and reconstructed data of each edge. PiCOL-3 achieves smaller MAEs on Edge 2-1 (where incident happens), and its neighboring edges 1-2, 1-70, 2-10, 3-1, 71-1, and 9-2.  }
    \label{tab:link-mae}
\end{table}

\subsection{Scalability of TTC-X}
{In urban transportation management, scalability is a crucial factor when deploying the TTC-X system. The foundation of TTC-X lies in the mobilized power of widely installed PTCs, enabling cost-efficiency. Theoretically, TTC-X can be implemented and deployed at any traffic intersection with existing PTC infrastructure.}

{STGP employs an encoder-decoder structure with transformer multi-head attention and graph convolution, enabling adaptation to large-scale urban road networks. Multi-head global attention allows the model to capture long-range dependencies in traffic data, which are crucial when scaling to extensive networks where distant road segments might influence each other due to common commuting patterns or incidents. Meanwhile, graph convolution in STGP effectively manage local spatial dependencies, aggregating information from neighboring nodes to accurately model local traffic dynamics. Although the experiments utilized a relatively small network, the dual approach of STGP is well-suited for providing traffic state estimation to larger urban areas.}

{Moreover, the PiCOL controller, powered by an online learning algorithm, does not require pre-training or fine-tuning. Consequently, PiCOL enjoys plug-and-play operation and is readily available for deployment in large networks without any prior knowledge or training. In addition, PiCOL is a distributed control mechanism where each camera simultaneously updates its control policy in coordination with others. The local policy update at each camera reduces computation overhead and is scalable to large networks. As one can see from Eq.\eqref{eq:exp3}, each camera's control computation complexity grows linearly with respect to its total number of actions (i.e., the number of edges) when evaluating the correlated exponential weights. For example, in the transportation network environment, one key intersection usually conjuncts 4 or 5 road segments, resulting in modest computation expenditure.}

\section{Implication}
\paragraph{Connection with the Cyber-Physical World}
The deployment of the Traffic-responsive Tilt Camera surveillance system (TTC-X) within a cyber-physical framework presents a strategic alternative to direct real-world implementation, primarily due to challenges in safety and cost. Implementing such advanced systems directly in real-world settings often involves high costs and potential safety risks, especially during the initial deployment phases. For example, adjusting traffic control in real-time based on system recommendations without thorough testing could lead to unpredictable scenarios, potentially endangering public safety and requiring significant financial investment for widespread installation and maintenance.

\begin{figure}[t!]
    \centering
    \includegraphics[width=1\columnwidth]{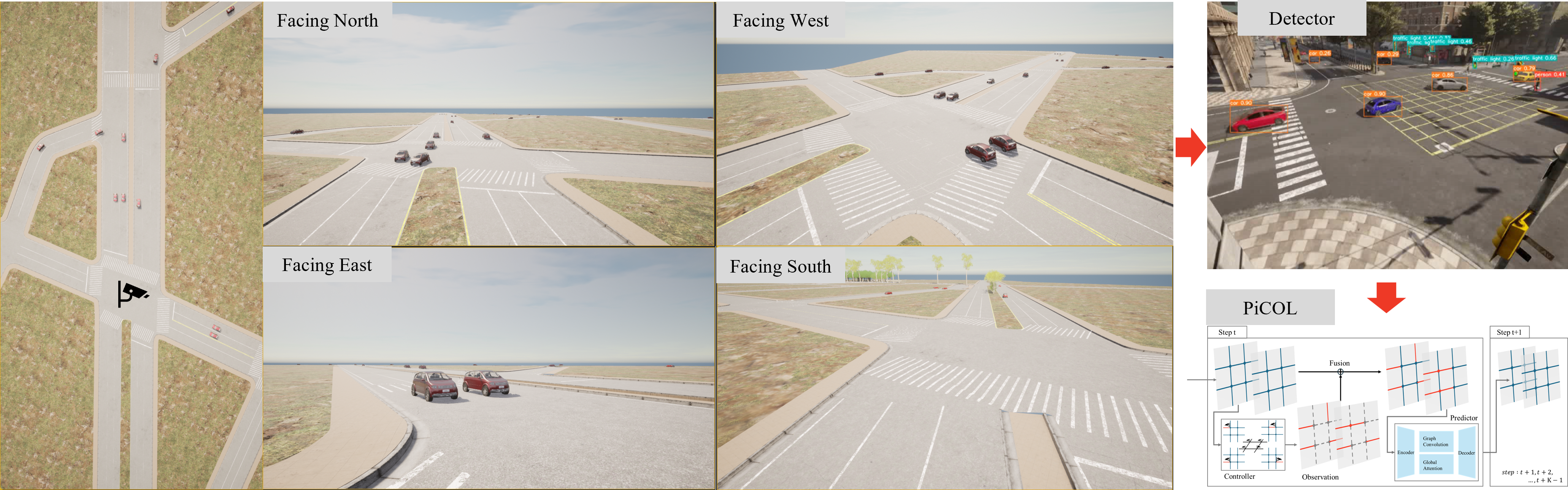}
    \caption{Cyber-Physical world implementation using CARLA-SUMO Co-simulation framework based on Flatbush corridor's network.}
    \label{fig:carla}
\end{figure}

To mitigate these challenges, the cyber-physical world serves as a bridge, allowing for the collecting and processing of real-world data in a controlled, virtual environment. This approach provides decision-supporting information or direct guidance to transportation agencies without the immediate risks and costs associated with physical deployment. The CARLA-SUMO co-simulation system is a prime example of this implementation, which creates a digital twin of the real-world traffic environment \cite{Dosovitskiy17}. This system allows for the testing and refining of the TTC-X in a virtual setup that closely mirrors actual traffic conditions. Figure \ref{fig:carla} shows a sample of the cyber-physical world and the connection to the proposed framework (Detector + PiCOL).

In the CARLA-SUMO co-simulation, various scenarios, such as network-level traffic flow optimization, route-level dynamic planning, and link-level traffic state forecasting, can be replicated and analyzed. This digital twin setup provides a safe and cost-effective platform for rigorous testing and optimization of the TTC-X framework, ensuring its readiness and reliability for eventual real-world deployment.

\paragraph{Connection with the Real World}
The TTC-X is designed as a plug-and-play system for real-world application, emphasizing ease of deployment and cost-effectiveness. This approach ensures the system can be integrated into existing traffic management infrastructures with minimal disruption and additional expense. The system's flexibility allows it to be tailored to specific urban traffic scenarios, making it a versatile tool for a wide range of traffic management tasks.

{The TTC-X can be rapidly deployed to new traffic intersections using their existing camera infrastructure, enabling immediate expansion to new surveillance and management neighborhoods within the road network. Additionally, the TTC-X is compatible with other existing traffic sensing infrastructures, such as loop detectors along road segments. This capability highlights the key advantage of using real-time fusion of multi-source traffic information.}

{One practical application is to feed real-time sensing data from loop detectors into the TTC-X predictor. This integration will enhance the predictability of the STGP by providing additional data points. Another approach involves using the TTC-X controller to enable multi-task surveillance and sensing simultaneously. Since loop detectors can supply real-time data for specific road segments, the PTCs in TTC-X can adjust to focus on other areas without loop detectors. This cooperative strategy allows for capturing network-level, route-level, and link-level information concurrently, optimizing the overall traffic management system.}

\section{Conclusion}

This study presents the Traffic-responsive Tilt Camera surveillance system (TTC-X), a comprehensive framework utilizing advanced machine learning for urban traffic management. The proposed system incorporates a novel Predictive Correlated Online Learning (PiCOL) module featuring the integration of an online learning controller and Spatial-Temporal Graph Predictor (STGP), enhancing real-time traffic estimation and camera control. Experiments were conducted in a simulated environment, which was calibrated using real-world traffic data from Brooklyn, New York, focusing on scenarios like maximum traffic flow capture, dynamic route planning, and link-level traffic state reconstruction and forecasting. TTC-X demonstrated significant improvements in capturing vehicles, adjusting optimal routes during incidents, and reconstructing traffic states. These scenarios demonstrated the capability of TTC-X in terms of its cost-effectiveness, rapid responsiveness to variant traffic situations as well as accurate capture of traffic states, showcasing its potential for effective traffic management in diverse urban settings. 

Future work should consider expanding the system's adaptability by refining PiCOL for more real-world dynamic traffic scenarios and improving the PTC control with consideration of decentralization approaches, such as game-theoretic learning algorithms. Additionally, given the good scalability of TTC-X, we will consider adding more traffic objects such as pedestrians, cyclists, and infrastructure to showcase more interactions between different traffic objects such as vehicle-to-pedestrian, and vehicle-to-infrastructure applications.

\section{Acknowledgements}
This work was supported by the C2SMARTER, a Tier 1 U.S. Department of Transportation (USDOT) funded University Transportation Center (UTC) led by New York University funded by USDOT. The contents of this paper only reflect the views of the authors who are responsible for the facts and do not represent any official views of any sponsoring organizations or agencies.

\section{Author Contributions}
The authors confirm their contribution to the paper as follows. Conceptualization: Zilin Bian, Tao Li, Fan Zuo; Data curation: Fan Zuo and Haozhe Lei; Formal analysis: Tao Li, Zilin Bian, Haozhe Lei; Methodology: Tao Li, Zilin Bian, Haozhe Lei, Fan Zuo, Ya-Ting Yang; Validation: Tao Li, Zilin Bian, Haozhe Lei; Visualization: Ya-Ting Yang, Fan Zuo, Tao Li, Zilin Bian; Writing - original draft: Tao Li, Zilin Bian, Fan Zuo, Ya-Ting Yang; and Writing - review \& editing: Tao Li, Zilin Bian, Fan Zuo, Kaan Ozbay, Quanyan Zhu, Zhenning Li. All authors reviewed the results and approved the final version of the manuscript.

\bibliography{ref} 

\begin{thebibliography}{10}

\bibitem{bian2019estimating}
Zilin Bian and Kaan Ozbay.
\newblock Estimating uncertainty of work zone capacity using neural network models.
\newblock {\em Transportation research record}, 2673(2):49--59, 2019.

\bibitem{bian2021time}
Zilin Bian, Fan Zuo, Jingqin Gao, Yanyan Chen, Sai Sarath Chandra~Pavuluri Venkata, Suzana~Duran Bernardes, Kaan Ozbay, Xuegang~Jeff Ban, and Jingxing Wang.
\newblock Time lag effects of covid-19 policies on transportation systems: A comparative study of new york city and seattle.
\newblock {\em Transportation Research Part A: Policy and Practice}, 145:269--283, 2021.

\bibitem{akilan2019video}
Thangarajah Akilan, QM~Jonathan Wu, and Wandong Zhang.
\newblock Video foreground extraction using multi-view receptive field and encoder--decoder dcnn for traffic and surveillance applications.
\newblock {\em IEEE Transactions on Vehicular Technology}, 68(10):9478--9493, 2019.

\bibitem{haghighat2023computer}
Arya Haghighat and Anuj Sharma.
\newblock A computer vision-based deep learning model to detect wrong-way driving using pan--tilt--zoom traffic cameras.
\newblock {\em Computer-Aided Civil and Infrastructure Engineering}, 38(1):119--132, 2023.

\bibitem{yang2023cooperative}
Hao~Frank Yang, Jiarui Cai, Chenxi Liu, Ruimin Ke, and Yinhai Wang.
\newblock Cooperative multi-camera vehicle tracking and traffic surveillance with edge artificial intelligence and representation learning.
\newblock {\em Transportation research part C: emerging technologies}, 148:103982, 2023.

\bibitem{redmon2016look}
Joseph Redmon, Santosh Divvala, Ross Girshick, and Ali Farhadi.
\newblock You only look once: Unified, real-time object detection, 2016.

\bibitem{wojke2017simple}
Nicolai Wojke, Alex Bewley, and Dietrich Paulus.
\newblock Simple online and realtime tracking with a deep association metric, 2017.

\bibitem{Tang_2019}
Zheng Tang and Jenq-Neng Hwang.
\newblock Moana: An online learned adaptive appearance model for robust multiple object tracking in 3d.
\newblock {\em IEEE Access}, 7:31934–31945, 2019.

\bibitem{neupane2022real}
Bipul Neupane, Teerayut Horanont, and Jagannath Aryal.
\newblock Real-time vehicle classification and tracking using a transfer learning-improved deep learning network.
\newblock {\em Sensors}, 22(10):3813, 2022.

\bibitem{shen2023triplet}
Fei Shen, Xiaoyu Du, Liyan Zhang, and Jinhui Tang.
\newblock Triplet contrastive learning for unsupervised vehicle re-identification.
\newblock {\em arXiv preprint arXiv:2301.09498}, 2023.

\bibitem{zhou2022gan}
Zhili Zhou, Yujiang Li, Jin Li, Keping Yu, Guang Kou, Meimin Wang, and Brij~Bhooshan Gupta.
\newblock Gan-siamese network for cross-domain vehicle re-identification in intelligent transport systems.
\newblock {\em IEEE Transactions on Network Science and Engineering}, 2022.

\bibitem{oladimeji2023smart}
Damilola Oladimeji, Khushi Gupta, Nuri~Alperen Kose, Kubra Gundogan, Linqiang Ge, and Fan Liang.
\newblock Smart transportation: an overview of technologies and applications.
\newblock {\em Sensors}, 23(8):3880, 2023.

\bibitem{aissaoui2014advanced}
Raik Aissaoui, Hamid Menouar, Amine Dhraief, Fethi Filali, Abdelfettah Belghith, and Adnan Abu-Dayya.
\newblock Advanced real-time traffic monitoring system based on v2x communications.
\newblock In {\em 2014 IEEE International Conference on Communications (ICC)}, pages 2713--2718. IEEE, 2014.

\bibitem{khaliq2019road}
Kishwer~Abdul Khaliq, Omer Chughtai, Abdullah Shahwani, Amir Qayyum, and J{\"u}rgen Pannek.
\newblock Road accidents detection, data collection and data analysis using v2x communication and edge/cloud computing.
\newblock {\em Electronics}, 8(8):896, 2019.

\bibitem{li2020trajectory}
Li~Li, Rui Jiang, Zhengbing He, Xiqun~Michael Chen, and Xuesong Zhou.
\newblock Trajectory data-based traffic flow studies: A revisit.
\newblock {\em Transportation Research Part C: Emerging Technologies}, 114:225--240, 2020.

\bibitem{sanguesa2015sensing}
Julio~A Sanguesa, Javier Barrachina, Manuel Fogue, Piedad Garrido, Francisco~J Martinez, Juan-Carlos Cano, Carlos~T Calafate, and Pietro Manzoni.
\newblock Sensing traffic density combining v2v and v2i wireless communications.
\newblock {\em Sensors}, 15(12):31794--31810, 2015.

\bibitem{rim2011estimation}
Heesub Rim, Cheol Oh, Kyungpyo Kang, and Seongho Kim.
\newblock Estimation of lane-level travel times in vehicle-to-vehicle and vehicle-to-infrastructure--based traffic information system.
\newblock {\em Transportation research record}, 2243(1):9--16, 2011.

\bibitem{han2020congestion}
Xiao Han, Guojiang Shen, Xi~Yang, and Xiangjie Kong.
\newblock Congestion recognition for hybrid urban road systems via digraph convolutional network.
\newblock {\em Transportation Research Part C: Emerging Technologies}, 121:102877, 2020.

\bibitem{wang2021adaptive}
Tong Wang, Jiahua Cao, and Azhar Hussain.
\newblock Adaptive traffic signal control for large-scale scenario with cooperative group-based multi-agent reinforcement learning.
\newblock {\em Transportation research part C: emerging technologies}, 125:103046, 2021.

\bibitem{song2019vision}
Huansheng Song, Haoxiang Liang, Huaiyu Li, Zhe Dai, and Xu~Yun.
\newblock Vision-based vehicle detection and counting system using deep learning in highway scenes.
\newblock {\em European Transport Research Review}, 11(1):1--16, 2019.

\bibitem{zhang2019vehicle}
Fukai Zhang, Ce~Li, and Feng Yang.
\newblock Vehicle detection in urban traffic surveillance images based on convolutional neural networks with feature concatenation.
\newblock {\em Sensors}, 19(3):594, 2019.

\bibitem{khalifa2020novel}
Anouar~Ben Khalifa, Ihsen Alouani, Mohamed~Ali Mahjoub, and Atika Rivenq.
\newblock A novel multi-view pedestrian detection database for collaborative intelligent transportation systems.
\newblock {\em Future generation computer systems}, 113:506--527, 2020.

\bibitem{el2020pedestrian}
Sara El~Hamdani, Nabil Benamar, and Mohamed Younis.
\newblock Pedestrian support in intelligent transportation systems: challenges, solutions and open issues.
\newblock {\em Transportation research part C: emerging technologies}, 121:102856, 2020.

\bibitem{ferreira2022identifying}
Marta~Campos Ferreira, Paulo~Dias Costa, Diogo Abrantes, Joana Hora, Soraia Fel{\'\i}cio, Miguel Coimbra, and Teresa~Galv{\~a}o Dias.
\newblock Identifying the determinants and understanding their effect on the perception of safety, security, and comfort by pedestrians and cyclists: A systematic review.
\newblock {\em Transportation research part F: traffic psychology and behaviour}, 91:136--163, 2022.

\bibitem{yu2020study}
Xiaoyan Yu and Marin Marinov.
\newblock A study on recent developments and issues with obstacle detection systems for automated vehicles.
\newblock {\em Sustainability}, 12(8):3281, 2020.

\bibitem{6875912}
Bin Tian, Brendan~Tran Morris, Ming Tang, Yuqiang Liu, Yanjie Yao, Chao Gou, Dayong Shen, and Shaohu Tang.
\newblock Hierarchical and networked vehicle surveillance in its: A survey.
\newblock {\em IEEE Transactions on Intelligent Transportation Systems}, 16(2):557--580, 2015.

\bibitem{bommes2016video}
Michael Bommes, Adrian Fazekas, Tobias Volkenhoff, and Markus Oeser.
\newblock Video based intelligent transportation systems--state of the art and future development.
\newblock {\em Transportation Research Procedia}, 14:4495--4504, 2016.

\bibitem{yang2018vehicle}
Zi~Yang and Lilian~SC Pun-Cheng.
\newblock Vehicle detection in intelligent transportation systems and its applications under varying environments: A review.
\newblock {\em Image and Vision Computing}, 69:143--154, 2018.

\bibitem{7458203}
Sokèmi René~Emmanuel Datondji, Yohan Dupuis, Peggy Subirats, and Pascal Vasseur.
\newblock A survey of vision-based traffic monitoring of road intersections.
\newblock {\em IEEE Transactions on Intelligent Transportation Systems}, 17(10):2681--2698, 2016.

\bibitem{koller1993model}
Dieter Koller, Kostas Daniilidis, and Hans-Hellmut Nagel.
\newblock Model-based object tracking in monocular image sequences of road traffic scenes.
\newblock {\em International Journal of Computer 11263on}, 10:257--281, 1993.

\bibitem{unzueta2011adaptive}
Luis Unzueta, Marcos Nieto, Andoni Cort{\'e}s, Javier Barandiaran, Oihana Otaegui, and Pedro S{\'a}nchez.
\newblock Adaptive multicue background subtraction for robust vehicle counting and classification.
\newblock {\em IEEE Transactions on Intelligent Transportation Systems}, 13(2):527--540, 2011.

\bibitem{morris2008learning}
Brendan~Tran Morris and Mohan~Manubhai Trivedi.
\newblock Learning, modeling, and classification of vehicle track patterns from live video.
\newblock {\em IEEE transactions on intelligent transportation systems}, 9(3):425--437, 2008.

\bibitem{9714212}
Xingchen Zhang, Yuxiang Feng, Panagiotis Angeloudis, and Yiannis Demiris.
\newblock Monocular visual traffic surveillance: A review.
\newblock {\em IEEE Transactions on Intelligent Transportation Systems}, 23(9):14148--14165, 2022.

\bibitem{fei2023multi}
Lunlin Fei and Bing Han.
\newblock Multi-object multi-camera tracking based on deep learning for intelligent transportation: a review.
\newblock {\em Sensors}, 23(8):3852, 2023.

\bibitem{yu2020forecasting}
Byeonghyeop Yu, Yongjin Lee, and Keemin Sohn.
\newblock Forecasting road traffic speeds by considering area-wide spatio-temporal dependencies based on a graph convolutional neural network (gcn).
\newblock {\em Transportation research part C: emerging technologies}, 114:189--204, 2020.

\bibitem{zhang2019spatial}
Chenhan Zhang, JQ~James, and Yi~Liu.
\newblock Spatial-temporal graph attention networks: A deep learning approach for traffic forecasting.
\newblock {\em IEEE Access}, 7:166246--166256, 2019.

\bibitem{do2019effective}
Loan~NN Do, Hai~L Vu, Bao~Q Vo, Zhiyuan Liu, and Dinh Phung.
\newblock An effective spatial-temporal attention based neural network for traffic flow prediction.
\newblock {\em Transportation research part C: emerging technologies}, 108:12--28, 2019.

\bibitem{vaswani2017attention}
Ashish Vaswani, Noam Shazeer, Niki Parmar, Jakob Uszkoreit, Llion Jones, Aidan~N Gomez, {\L}ukasz Kaiser, and Illia Polosukhin.
\newblock Attention is all you need.
\newblock {\em Advances in neural information processing systems}, 30, 2017.

\bibitem{huang2022dynamical}
Feihu Huang, Peiyu Yi, Jince Wang, Mengshi Li, Jian Peng, and Xi~Xiong.
\newblock A dynamical spatial-temporal graph neural network for traffic demand prediction.
\newblock {\em Information Sciences}, 594:286--304, 2022.

\bibitem{zhang2022trajectory}
Kunpeng Zhang, Xiaoliang Feng, Lan Wu, and Zhengbing He.
\newblock Trajectory prediction for autonomous driving using spatial-temporal graph attention transformer.
\newblock {\em IEEE Transactions on Intelligent Transportation Systems}, 23(11):22343--22353, 2022.

\bibitem{shai_online}
Shai Shalev-Shwartz.
\newblock {Online Learning and Online Convex Optimization}.
\newblock {\em Foundations and Trends® in Machine Learning}, 4(2):107--194, 2011.

\bibitem{slivkins19bandit}
Aleksandrs Slivkins.
\newblock {Introduction to Multi-Armed Bandits}.
\newblock {\em Foundations and Trends® in Machine Learning}, 12(1-2):1--286, 2019.

\bibitem{schapire99ew}
Yoav Freund and Robert~E. Schapire.
\newblock {Adaptive Game Playing Using Multiplicative Weights}.
\newblock {\em Games and Economic Behavior}, 29(1-2):79--103, 1999.

\bibitem{bianchi02adv-bandit}
Peter Auer, Nicol Cesa-Bianchi, Yoav Freund, and Robert~E. Schapire.
\newblock {The Nonstochastic Multiarmed Bandit Problem}.
\newblock {\em SIAM Journal on Computing}, 32(1):48--77, 2002.

\bibitem{Rakhlin13pred}
Alexander Rakhlin and Karthik Sridharan.
\newblock {Online Learning with Predictable Sequences}.
\newblock 30:993--1019, 2013.

\bibitem{jadbabaie15pred_opt}
Ali Jadbabaie, Alexander Rakhlin, Shahin Shahrampour, and Karthik Sridharan.
\newblock {Online Optimization : Competing with Dynamic Comparators}.
\newblock In Guy Lebanon and S.~V.~N. Vishwanathan, editors, {\em Proceedings of the Eighteenth International Conference on Artificial Intelligence and Statistics}, volume~38 of {\em Proceedings of Machine Learning Research}, pages 398--406, San Diego, California, USA, 09--12 May 2015. PMLR.

\bibitem{tao_info}
Tao Li, Yuhan Zhao, and Quanyan Zhu.
\newblock {The role of information structures in game-theoretic multi-agent learning}.
\newblock {\em Annual Reviews in Control}, 53:296--314, 2022.

\bibitem{Bianchi_Lugosi_2006}
Nicolo Cesa-Bianchi and Gabor Lugosi.
\newblock {\em Prediction, Learning, and Games}.
\newblock Cambridge University Press, 2006.

\bibitem{hart00regret_match}
Sergiu Hart and Andreu Mas-Colell.
\newblock {A Simple Adaptive Procedure Leading to Correlated Equilibrium}.
\newblock {\em Econometrica}, 68(5):1127--1150, 2000.
\newblock regret matching.

\bibitem{perchet14blackwell}
Vianney Perchet.
\newblock {Approachability, regret and calibration: Implications and equivalences}.
\newblock {\em Journal of Dynamics \& Games}, 1(2):181--254, 2014.

\bibitem{hong23signal-control}
Wanshi Hong, Gang Tao, Hong Wang, and Chieh Wang.
\newblock {Traffic Signal Control With Adaptive Online-Learning Scheme Using Multiple-Model Neural Networks}.
\newblock {\em IEEE Transactions on Neural Networks and Learning Systems}, 34(10):7838--7850, 2023.

\bibitem{guo21traffic-forecast}
Zhengang Guo, Yingfeng Zhang, Jingxiang Lv, Yang Liu, and Ying Liu.
\newblock {An Online Learning Collaborative Method for Traffic Forecasting and Routing Optimization}.
\newblock {\em IEEE Transactions on Intelligent Transportation Systems}, 22(10):6634--6645, 2021.

\bibitem{zuo2020interactive}
Fan Zuo, Jingxing Wang, Jingqin Gao, Kaan Ozbay, Xuegang~Jeff Ban, Yubin Shen, Hong Yang, and Shri Iyer.
\newblock An interactive data visualization and analytics tool to evaluate mobility and sociability trends during covid-19.
\newblock {\em arXiv preprint arXiv:2006.14882}, 2020.

\bibitem{zuo2021reference}
Fan Zuo, Jingqin Gao, Abdullah Kurkcu, Hong Yang, Kaan Ozbay, and Qingyu Ma.
\newblock Reference-free video-to-real distance approximation-based urban social distancing analytics amid covid-19 pandemic.
\newblock {\em Journal of Transport \& Health}, 21:101032, 2021.

\bibitem{bochkovskiy2020yolov4}
Alexey Bochkovskiy, Chien-Yao Wang, and Hong-Yuan~Mark Liao.
\newblock Yolov4: Optimal speed and accuracy of object detection.
\newblock {\em arXiv preprint arXiv:2004.10934}, 2020.

\bibitem{du2023strongsort}
Yunhao Du, Zhicheng Zhao, Yang Song, Yanyun Zhao, Fei Su, Tao Gong, and Hongying Meng.
\newblock Strongsort: Make deepsort great again.
\newblock {\em IEEE Transactions on Multimedia}, 2023.

\bibitem{bewley2016simple}
Alex Bewley, Zongyuan Ge, Lionel Ott, Fabio Ramos, and Ben Upcroft.
\newblock Simple online and realtime tracking.
\newblock In {\em 2016 IEEE international conference on image processing (ICIP)}, pages 3464--3468. IEEE, 2016.

\bibitem{bertsekas2011dynamic}
Dimitri~P Bertsekas et~al.
\newblock Dynamic programming and optimal control 3rd edition, volume ii.
\newblock {\em Belmont, MA: Athena Scientific}, 1, 2011.

\bibitem{zuo2020microscopic}
F~Zuo, K~Ozbay, A~Kurkcu, J~Gao, Huajie Yang, and K~Xie.
\newblock Microscopic simulation based study of pedestrian safety applications at signalized urban crossings in a connected-automated vehicle environment and reinforcement learning based optimization of vehicle decisions.
\newblock {\em Advances in transportation studies}, 2020.

\bibitem{tao_multiRL}
Tao Li and Quanyan Zhu.
\newblock {On Convergence Rate of Adaptive Multiscale Value Function Approximation for Reinforcement Learning}.
\newblock {\em 2019 IEEE 29th International Workshop on Machine Learning for Signal Processing (MLSP)}, pages 1--6, 2019.

\bibitem{sutton_PG}
Richard~S Sutton, David~A McAllester, Satinder~P Singh, and Yishay Mansour.
\newblock {Policy gradient methods for reinforcement learning with function approximation}.
\newblock In {\em Advances in Neural Information Processing Systems 12}, pages 1057---1063. MIT press, 2000.

\bibitem{tao22confluence}
Tao Li, Guanze Peng, Quanyan Zhu, and Tamer Baar.
\newblock {The Confluence of Networks, Games, and Learning a Game-Theoretic Framework for Multiagent Decision Making Over Networks}.
\newblock {\em IEEE Control Systems}, 42(4):35--67, 2022.

\bibitem{levine2020offline}
Sergey Levine, Aviral Kumar, George Tucker, and Justin Fu.
\newblock Offline reinforcement learning: Tutorial, review, and perspectives on open problems.
\newblock {\em arXiv preprint arXiv:2005.01643}, 2020.

\bibitem{finn2019online}
Chelsea Finn, Aravind Rajeswaran, Sham Kakade, and Sergey Levine.
\newblock Online meta-learning.
\newblock In {\em International Conference on Machine Learning}, pages 1920--1930. PMLR, 2019.

\bibitem{tao23cola}
Tao Li, Haozhe Lei, and Quanyan Zhu.
\newblock {Self-Adaptive Driving in Nonstationary Environments through Conjectural Online Lookahead Adaptation}.
\newblock {\em 2023 IEEE International Conference on Robotics and Automation (ICRA)}, 00:7205--7211, 2023.

\bibitem{pan2023meta-sg}
Yunian Pan, Tao Li, Henger Li, Tianyi Xu, Zizhan Zheng, and Quanyan Zhu.
\newblock A first order meta stackelberg method for robust federated learning.
\newblock {\em The 2nd Workshop on New Frontiers In Adversarial Machine Learning at 40th International Conference on Machine Learning}, 2023.

\bibitem{Tao_blackwell}
Tao Li, Guanze Peng, and Quanyan Zhu.
\newblock {Blackwell Online Learning for Markov Decision Processes}.
\newblock {\em 2021 55th Annual Conference on Information Sciences and Systems (CISS)}, 00:1--6, 2021.

\bibitem{robbins52mab}
Herbert Robbins.
\newblock {Some aspects of the sequential design of experiments}.
\newblock {\em Bulletin of the American Mathematical Society}, 58(5):527--535, 1952.

\bibitem{zhang2024semantic}
Kunpeng Zhang, Feng Zhou, Lan Wu, Na~Xie, and Zhengbing He.
\newblock Semantic understanding and prompt engineering for large-scale traffic data imputation.
\newblock {\em Information Fusion}, 102:102038, 2024.

\bibitem{guo2021learning}
Shengnan Guo, Youfang Lin, Huaiyu Wan, Xiucheng Li, and Gao Cong.
\newblock Learning dynamics and heterogeneity of spatial-temporal graph data for traffic forecasting.
\newblock {\em IEEE Transactions on Knowledge and Data Engineering}, 34(11):5415--5428, 2021.

\bibitem{kipf2016semi}
Thomas~N Kipf and Max Welling.
\newblock Semi-supervised classification with graph convolutional networks.
\newblock {\em arXiv preprint arXiv:1609.02907}, 2016.

\bibitem{SUMO2018}
Pablo~Alvarez Lopez, Michael Behrisch, Laura Bieker-Walz, Jakob Erdmann, Yun-Pang Fl{\"o}tter{\"o}d, Robert Hilbrich, Leonhard L{\"u}cken, Johannes Rummel, Peter Wagner, and Evamarie Wie{\ss}ner.
\newblock Microscopic traffic simulation using sumo.
\newblock In {\em The 21st IEEE International Conference on Intelligent Transportation Systems}. IEEE, 2018.

\bibitem{paszke2019pytorch}
Adam Paszke, Sam Gross, Francisco Massa, Adam Lerer, James Bradbury, Gregory Chanan, Trevor Killeen, Zeming Lin, Natalia Gimelshein, Luca Antiga, et~al.
\newblock Pytorch: An imperative style, high-performance deep learning library.
\newblock {\em Advances in neural information processing systems}, 32, 2019.

\bibitem{Dosovitskiy17}
Alexey Dosovitskiy, German Ros, Felipe Codevilla, Antonio Lopez, and Vladlen Koltun.
\newblock {CARLA}: {An} open urban driving simulator.
\newblock In {\em Proceedings of the 1st Annual Conference on Robot Learning}, pages 1--16, 2017.

\end{thebibliography}
\bibliographystyle{unsrt}
\end{document}